\pdfoutput=1
\documentclass{article}
\usepackage[margin=1.0in]{geometry}

\usepackage{cite}
\usepackage{amsmath,amssymb,amsfonts}
\usepackage{algorithmic}
\usepackage{graphicx}
\usepackage{textcomp}
\usepackage{mathtools}

\usepackage{helvet}

\usepackage{mathtools}
\usepackage{cite}
\usepackage{appendix}
\usepackage[affil-it]{authblk} 

\newcommand{\reffig}[1] {\ref{#1}}
\newcommand{\refeqn}[1] {(\ref{#1})}
\newcommand{\reftable}[1] {\ref{#1}}

\newcommand{\dee}{{\rm d}}
\newcommand{\R}{\mathbb{R}}

\newcommand{\Norm}[1]{\left\Vert#1\right\Vert}
\newcommand{\muaa}{\mu_{\mathrm{a}}}
\newcommand{\muas}{\mu_{\mathrm{s}}}
\newcommand{\gruneisen}[1]{G}

\newcommand{\revold}[1]{}
\newcommand{\revnew}[1]{{#1}}

\newcommand\blfootnote[1]{%
  \begingroup
  \renewcommand\thefootnote{}\footnote{#1}%
  \addtocounter{footnote}{-1}%
  \endgroup
}

\begin{document}

\title{Perturbation Monte Carlo Method for \\ Quantitative Photoacoustic Tomography}

%\author{Aleksi~Leino,
%        Tuomas~Lunttila,
%        Meghdoot~Mozumder,
%        Aki~Pulkkinen,	
%        and~Tanja~Tarvainen}

\author[a]{Aleksi Leino}
\author[a]{Tuomas Lunttila}
\author[a]{Meghdoot Mozumder}
\author[a]{Aki Pulkkinen}
\author[a,b]{Tanja Tarvainen}

\affil[a]{\small Department of Applied Physics, University of Eastern Finland, P.O. Box 1627, 70211 Kuopio, Finland}
\affil[b]{Department of Computer Science, University College London, Gower Street, WC1E 6BT, London, United Kingdom}

\maketitle

%=========================================================
\begin{abstract}
  Quantitative photoacoustic tomography aims at estimating optical parameters from photoacoustic images that are formed utilizing the photoacoustic effect caused by the absorption of an externally introduced light pulse. 
  This optical parameter estimation is an ill-posed inverse problem, and thus it is sensitive to measurement and modeling errors. 
  In this work, we propose a novel way to solve the inverse problem of quantitative photoacoustic tomography based on the perturbation Monte Carlo method. 
  Monte Carlo method for light propagation is a stochastic approach for simulating photon trajectories in a medium with scattering particles.
  It is widely accepted as an accurate method to simulate light propagation in tissues.
  Furthermore, it is numerically  robust and easy to implement. 
  Perturbation Monte Carlo maintains this robustness and enables forming gradients for the solution of the inverse problem.
  We validate the method and apply it in the framework of Bayesian inverse problems. 
  The simulations show that the perturbation Monte Carlo method can be used to estimate spatial distributions of both absorption and scattering parameters simultaneously.
  These estimates are qualitatively good and quantitatively accurate also in parameter scales that are realistic for biological tissues.
\end{abstract}

\blfootnote{\textcopyright \phantom{c}2020 IEEE.  Personal use of this material is permitted.  Permission from IEEE must be obtained for all other uses, in any current or future media, including reprinting/republishing this material for advertising or promotional purposes, creating new collective works, for resale or redistribution to servers or lists, or reuse of any copyrighted component of this work in other works.}

%Keywords: Quantitative photoacoustic tomography, perturbation Monte Carlo, inverse problems, photoacoustic imaging, optoacoustic imaging, image reconstruction  

Photoacoustic tomography (PAT) is an imaging modality
based on the photoacoustic effect generated by the absorption of an
externally introduced light pulse in the imaged target.  PAT combines
optical contrast and specificity with the high spatial resolution of
ultrasound.  It has various applications in imaging of soft biological
tissue, such as imaging human blood vessels, microvasculature of
tumors and the cerebral cortex in small animals \cite{xu2006, li2009,
  wang2009, beard2011, xia2014,wang2016,weber2016,brunker2017}.
Quantitative photoacoustic tomography (QPAT) continues from the
conventional photoacoustic images and aims at estimating the spatial
distributions of the optical parameters \cite{cox2012a}.  This optical
inverse problem of QPAT is ill-posed, meaning that even small errors
in measurements or modeling can lead to large errors in the solution.
Therefore, solution of the QPAT inverse problem relies strongly on
accurate modeling of light transport.

A widely accepted forward model for light propagation in scattering
medium such as biological tissue is the radiative transfer equation
(RTE) \cite{ishimaru78a,arridge99}.  Given the light source, geometry
and the optical parameters of the medium, it can solve the light
fluence and optical energy absorption into the tissue.  Due to the
computational complexity of the RTE, its approximations, such as the
diffusion approximation, are generally applied in optical imaging
\cite{arridge99}.  However, the diffusion approximation is not valid
in typical QPAT imaging situations where the size of the imaged
targets corresponds approximately to a few scattering lengths.
Alternatively to the deterministic models, the Monte Carlo method can
be used to simulate light propagation in tissue.  Monte Carlo is a
stochastic method that can be used to simulate light tissue
interactions.  It has been widely utilized in biomedical optics, see
e.g. \cite{prahl89,wang1995,sassaroli2012,zhu2013,hayakawa2014}, and
various open-source Monte Carlo implementations have been published
\cite{wang1995,fang2009,yang2013,cassidy2018,liu2015,Leino19}\revnew{.}

The optical inverse problem of QPAT is typically formulated as a
minimization problem that is solved using methods of numerical
optimization \cite{cox2012a}.  Essentially, one minimizes the
difference of the 'measured' optical energy density and that produced
by a numerical solution of a forward model.  This is a large scale
inverse problem with a large number of both unknown parameters and
data points.  Although methods for utilizing the RTE in the optical
inverse problem have been presented
\cite{tarvainen2012,saratoon2013,mamonov2014,haltmeier2015}, the
numerical implementations still suffer from the computationally
expensive nature of the problem.

In this work, we use the Monte Carlo method in the solution of the
optical inverse problem of QPAT.  Previously Monte Carlo has been
utilized in QPAT inverse problem by assuming the scattering as known
and estimating the absorption \cite{buchmann2017,kaplan2017}.  In
practice, however, the scattering is not known and it needs to be
taken into account when solving the inverse problem.  Alternatively,
adjoint Monte Carlo models of radiance have been applied to formulate
the solution of the QPAT inverse problem 
% AKI: NOTE ISSUE WITH \REVOLD
% {\color{red}\cite{hochuli2016}}
\revnew{\cite{hochuli2016, Buchmann2019, buchmann2020}}.
This
approach necessitates solving the radiance in several points in space
and angular directions which causes challenges to storing the angular
solution and having good enough sampling to allow an acceptable level
of noise.  The approach has been utilized in QPAT in estimating either
absorption or scattering while keeping the other one as a known
constant 
% AKI: NOTE ISSUE WITH \REVOLD
% {\color{red}\cite{hochuli2016,Buchmann2019}.
\revnew{\cite{hochuli2016, Buchmann2019}}, \revnew{including a recent study with experimental data \cite{buchmann2020}.}

Herein, we introduce an approach to the QPAT inverse problem based on
the so-called perturbation Monte Carlo (PMC) concept
\cite{Ivan1991,Hayakawa2001}.  In PMC, the aim is to evaluate the
effect of a small change in the optical parameters, i.e. perturbation,
efficiently.  This is achieved by re-using the trajectories of photons
from an unperturbed simulation so that the trajectories do not need to
be re-generated for each perturbation \cite{Ivan1991}.  Previously,
PMC 
\revnew{and a similar so-called white Monte Carlo approach have}
\revold{has} 
been utilized in other optical tomographic imaging modalities
in diffuse optical tomography and fluorescence diffuse optical
tomography for example in
%\revold{\cite{Hayakawa2001,kumar2004,chen2009,sassaroli2011,zhang2012,yamamoto2016,yao2018}.}
\revnew{\cite{kienle1996,pifferi1998,Hayakawa2001,kumar2004, alerstam2008,chen2009,sassaroli2011,zhang2012,yamamoto2016,yao2018}.}
Compared to the adjoint Monte Carlo, PMC does not require forming the
radiance or its approximations, and thus it is numerically less
expensive.  We approach the QPAT inverse problem in the framework of
the Bayesian inverse problems, and estimate both absorption and
scattering simultaneously.  A methodology for forming the Jacobian for
the inverse problem of QPAT based on the PMC methodology is presented
and validated.  To our knowledge, this is the first work in which the
PMC methodology is formulated for QPAT, and the first study in which
the Monte Carlo method is used to estimate the spatial distributions
of absorption and scattering parameters simultaneously in QPAT.

The rest of the paper is organized so that the forward model and the
inverse problem of QPAT are described in Sec. \ref{sec:theory}.  
\revold{Monte Carlo is reviewed and the PMC methodology together formulation for Jacobian is described in Sec. \ref{sec:pmcpt}.}
\revnew{Monte Carlo is reviewed, PMC methodology is introduced, and formulation for Jacobians is provided in Sec. \ref{sec:pmcpt}.}
The methodology is evaluated with simulations and discussed in
Sec. \ref{sec:simulations} followed by conclusions in
Sec. \ref{sec:conclusions}.

%==========================================================================
% Section II: Forward and inverse problems of quantitative photoacoustic tomography
%==========================================================================

\section{Forward and inverse problems of quantitative photoacoustic tomography}
\label{sec:theory}

\subsection{Forward model for quantitative photoacoustic tomography}
\label{sec:forwardpt}

Modeling photoacoustic effect consists of modeling optical light propagation and acoustic ultrasound propagation together with their coupling. 
Due to the difference in time-scales between the absorption of light and propagation of ultrasound, the pressure increase due to the absorption of light can be regarded as instantaneous with respect to the acoustic model. 
Therefore, in the optical model, a time-independent light transport model can be used. 
Further, the coupling of the optical and acoustic models can be described by a linear model. 

\subsubsection{Optical model}

Let $r \in \R^d $ be a point located in a tissue region of
interest $\Omega  \subset \R^d$ with boundary $\partial \Omega$ where
$d= 2,3$ is the dimension of the domain, and and let $\hat{s} \in S^{d-1}$ 
denote a unit vector in the direction of interest. 
Light transport in biological tissue can be modeled with the radiative transfer equation 
\begin{equation}
\begin{dcases}
  \label{eq:rte_cw}
   \hat{s}
    \cdot \nabla \phi (r,\hat{s}) + (\muas(r) + \muaa(r)) \phi(r,\hat{s})  \\
    \qquad  = \, \muas (r)\int _{S^{d-1}}  \Theta (\hat{s} \cdot
    \hat{s}') \phi(r,\hat{s}') \dee \hat{s}', \qquad r \in \Omega \\
    \phi(r,{\hat s}) = \left \lbrace
    \begin{array}{cl} \phi _0(r,\hat{s}), & r \in  \xi
      _j, \quad \hat{s} \cdot {\hat n} < 0 \\ 
      0, & r \in \partial \Omega \backslash \xi_j,
      \quad  \hat{s} \cdot {\hat n} < 0. \end{array} \right.
\end{dcases}
\end{equation}
where $\phi(r,\hat{s})$ is the radiance, $\muas(r)$ is the scattering
coefficient, $\muaa(r)$ is the absorption coefficient, \mbox{$\Theta
  (\hat{s} \cdot \hat{s}')$} is the scattering phase function, 
\revold{$\phi_0(r,{\hat s},t)$}
\revnew{$\phi_0(r,{\hat s})$}
 is a boundary light source at a source position
\mbox{$\xi _j \subset \partial \Omega$} and $\hat{n}$ is an outward
unit normal \cite{ishimaru78a}.  The scattering phase function
\mbox{$\Theta (\hat{s} \cdot \hat{s}')$} describes the probability
that a photon with an initial direction ${\hat s}'$ will have a
direction ${\hat s}$ after a scattering event.  In optical imaging,
the most commonly applied phase function is the Henyey-Greenstein
scattering function \cite{henyey41} which is of the form
\begin{equation}
  \label{eq:henyeygreenstein2Dand3D}
  \Theta (\hat{s}\cdot \hat{s}') = \left \lbrace  \begin{array}{ll}  
 \frac{1}{2\pi } \frac{1-g^2}{\left( 1 + g^2 - 2g (\hat{s} \cdot \hat{s}') \right) }, &
 \quad d = 2 \\
    \frac{1}{4\pi } \frac{1-g^2}{\left(
      1 + g^2 -2g (\hat{s} \cdot \hat{s}') \right)^{3/2}}, & \quad d = 3
    \end{array} \right.
\end{equation}
where $g$ is the scattering anisotropy parameter that defines the shape of the
probability density. 
It \revold{gets} \revnew{has} values between \mbox{$-1 < g < 1$}, such that, 
if \mbox{$g=0$}, the scattering probability density is a
uniform distribution,  \mbox{$g > 0$} for forward dominated scattering, and  \mbox{$g < 0$} for backward dominated scattering. 

The total energy at position $r$, often called photon fluence
$\Phi(r)$, is obtained from \revnew{the} radiance by
\begin{equation}
  \label{eq:fluence}
  \Phi(r)  =\int _{S^{d-1}}\phi(r,\hat{s})\dee \hat{s}. 
\end{equation}
Further, the absorption of light creates an absorbed optical energy
density $H(r)$ given by
\begin{equation}
  \label{eq:absorbed_energy}
  H(r) = \muaa (r) \Phi (r).
\end{equation}
Here we approximate the solution of the RTE with the Monte Carlo
method as implemented in ValoMC software and MATLAB toolbox
\cite{Leino19}.

\subsubsection{Acoustic model}

Propagation of sound, created by the instantaneous photoacoustic effect, in an infinite domain composed of homogeneous non-attenuating medium is described by the acoustic initial value problem \cite{cox2005,li2009}
\begin{equation}
 \label{eq:wave_equation_pat}
 \begin{dcases}
   \frac{1}{v^2}\frac{\partial ^2 p(r,t)}{\partial t^2} - \nabla^2p(r,t) = 0, \qquad r \in \R^d  \\ p(r,t=0) = p_0(r) \\
   \frac{\partial}{\partial t} p(r,t=0) = 0 
 \end{dcases}
\end{equation}
where $p(r,t)$ is the acoustic pressure, $v$ is the speed of sound, $t$ is
the time, and $p_0(r)$ is the initial pressure distribution created by
the absorption of a light pulse.
The initial pressure is given by
\begin{equation}
\label{eq:initial_pressure}
p_0(r) = \left\lbrace \begin{array}{ll} G H(r), \quad & r \in \Omega \\ 0, \quad & r \in \R^d \backslash \Omega \end{array} \right.
\end{equation}
where $G$ is the Gr{\"u}neisen parameter for an absorbing fluid that is used to identify photoacoustic efficiency \cite{li2009}. 
Throughout this work, $G$ is treated as a known constant, although in general, this is not the case. 
The solution of the initial value problem is obtained by numerically approximating the solution of the wave equation using \emph{k}-space time-domain method implemented with the k-Wave MATLAB toolbox \cite{treeby2010a}. 

%=========================================================

\subsection{Inverse problem}
\label{sec:inverseproblem}

The inverse problem in QPAT is to solve the optical parameters in the medium when the measured pressure wave on the sensors and the amount of input light are given.  
This can be approached in one step by directly estimating the optical parameters from the photoacoustic time-series or in two steps by first considering the acoustic inverse problem and then the optical inverse problem. 
Here we take the two-step approach. Further, we concentrate on the optical inverse problem.

\subsubsection{Acoustic inverse problem}

In the acoustic inverse problem, the initial acoustic pressure distribution $p_0(r)$ is estimated from photoacoustic waves $p_S$ measured on the acoustic sensors outside the imaged target. 
We use a time-reversal method implemented with the k-Wave MATLAB
toolbox for the solution of the acoustic inverse problem
\cite{treeby2010a}.
In this approach, the recorded measurements $p_S(t)$ are used 
in time-reversed order as a time-varying Dirichlet boundary
condition at the sensor positions. 
The time evolution of the propagating acoustic wavefield imposed by
the Dirichlet boundary condition is calculated using the wave equation with zero initial conditions.  
The reconstructed initial pressure $\hat{p}_0$ is then obtained as an acoustic
pressure within the domain after a time $T$. 
The medium is assumed to be
non-absorbing and the speed of sound is assumed to be known.

\subsubsection{Optical inverse problem}

In the optical inverse problem of QPAT, the optical parameters of the
medium are estimated when the absorbed optical energy density (or the
initial pressure) and the input light illumination are given.
\revnew{Commonly multiple different illumination patterns are used to
  provide sufficient data enabling estimation of both the optical
  absorption and scattering.}
\revold{We} \revnew{Here, we} approach the problem in the framework of
Bayesian inverse problems \cite{kaipio05,tarvainen2013}
\revold{.}\revnew{, while the PMC method could be implemented in other
  frameworks as well}.

A discrete observation model for QPAT in the presence of additive
noise is 
\revold{$H_{\rm meas} = H(r) + e$}
\revnew{
  \begin{equation}
    \label{eq:observation_model}
    H_{\rm meas} = H(x) + e
  \end{equation}
}
where $H_{\rm meas} \in \R^{m}$ is a data vector where $m$ is the
number of data which in the case of QPAT is the number of
illuminations multiplied with the number of discretization points to
represent the data space, 
\revnew{$x = \begin{pmatrix}
\mu_\mathrm{a} \\ \mu_\mathrm{s}
\end{pmatrix}$ are the optical parameters of interest, with}
\revold{$\mu_\mathrm{a} = (\mu_{{\rm a}_1},\ldots ,\mu_{{\rm a}_K})^{\rm T} \in \R^{K}$ and $\mu_\mathrm{s} = (\mu_{\rm s_{1}},\ldots ,\mu_{\rm s_{K}})^{\rm T}\in \R^{K}$}
\revnew{the absorption coefficients $\mu_\mathrm{a} = (\mu_{{\rm a}, 1},\ldots ,\mu_{{\rm a}, n})^{\rm T} \in \R^{n}$ and the scattering coefficients $\mu_\mathrm{s} = (\mu_{{\rm s}, 1},\ldots ,\mu_{{\rm s}, n})^{\rm T}\in \R^{n}$,}
\revold{are absorption and scattering coefficients,}\revold{$K$}\revnew{$n$} is the number of spatial discretization points,\revold{$H: \R^{2K} \mapsto\R^{m}$}
\revnew{$H: \R^{2n} \mapsto \R^{m}$} is the forward operator which maps the optical parameters to data
predictions, and $e \in \R^{m}$ denotes additive noise.

In the Bayesian approach to inverse problems, all parameters are
modeled as random variables.  Using Bayes' formula and following
derivation given e.g. in \cite{kaipio05}, the solution of the inverse
problem, i.e. the posterior distribution of the unknown parameters,
can be formulated.\revold{In principle, the posterior distribution could be estimated
  using Markov chain Monte Carlo (MCMC) methods.  However, these
  methods can be computationally too expensive in large dimensional
  tomographic inverse problems.  Therefore, point estimates such as
  the \emph{maximum a posteriori} (MAP) estimate are computed.}\revnew{In this work, we limit to evaluating the \emph{maximum a
    posteriori} (MAP) estimates based on the posterior distribution. }Thus, we seek to find absorption and scattering coefficients \revnew{by solving a minimization problem}
\revold{which
minimize $(\hat{\mu}_\mathrm{a},\hat{\mu}_\mathrm{s}) = \underset{(\mu_{\mathrm{a}},\mu_{\mathrm{s}})}{\arg\:\min} \:  \lbrace  \frac{1}{2}\Norm{ L_{e}(H_{\rm meas} - {H}(r)) - \eta_{e} }_2^2 +$}
\revold{$\frac{1}{2} \Norm{ L_{\mu_{\mathrm{a}}}(\mu_{\mathrm{a}} - \eta_{\mu_{\mathrm{a}}}) }_2^2 + \frac{1}{2} \Norm{ L_{\mu_{\mathrm{s}}}(\mu_{\mathrm{s}} - \eta_{\mu_{\mathrm{s}}}) }_2^2 \rbrace$}
\revnew{
  \begin{equation}
    \begin{split}
      \label{eq:MAP_QPAT}
      %(\hat{\mu} _\mathrm{a},\hat{\mu}_\mathrm{s}) & =  \underset{(\mu_{\mathrm{a}},\mu_{\mathrm{s}})}{\arg\:\min} \:  \left\lbrace  \frac{1}{2}\Norm{ L_{e}(H_{\rm meas} - {H}(x) - \eta_{e}) }_2^2 \right. \\ & \left. + \frac{1}{2} \Norm{ L_{\mu_{\mathrm{a}}}(\mu_{\mathrm{a}} - \eta_{\mu_{\mathrm{a}}}) }_2^2  + \frac{1}{2} \Norm{ L_{\mu_{\mathrm{s}}}(\mu_{\mathrm{s}} - \eta_{\mu_{\mathrm{s}}}) }_2^2 \right\rbrace
      (\hat{\mu} _\mathrm{a},\hat{\mu}_\mathrm{s}) & =  \underset{(\mu_{\mathrm{a}},\mu_{\mathrm{s}})}{\arg\:\min} \:  \left\lbrace  \frac{1}{2}\Norm{ \mathcal{L}_{e}(H_{\rm meas} - {H}(x) - \eta_{e}) }_2^2 \right. \\ & \left. + \frac{1}{2} \Norm{ \mathcal{L}_{\mu_{\mathrm{a}}}(\mu_{\mathrm{a}} - \eta_{\mu_{\mathrm{a}}}) }_2^2  + \frac{1}{2} \Norm{ \mathcal{L}_{\mu_{\mathrm{s}}}(\mu_{\mathrm{s}} - \eta_{\mu_{\mathrm{s}}}) }_2^2 \right\rbrace
    \end{split}
  \end{equation}
}
where noise is modeled as Gaussian with an expected value $\eta_{e}$
and 
\revold{$L_{e}$}
\revnew{$\mathcal{L}_{e}$}
being the Cholesky decomposition of the inverse of the
noise covariance matrix 
\revold{$\Gamma _{e}^{-1}=L_{e}^{\rm T}L_{e}$}
\revnew{$\Gamma_{e}^{-1} = \mathcal{L}_{e}^{\rm T} \mathcal{L}_{e}$}
\cite{tarvainen2013}.  Prior information of absorption and scattering
is described by Gaussian prior distributions where
$\eta_{\mu_{\mathrm{a}}}$ and $\eta_{\mu_{\mathrm{s}}}$ are the
expected values of absorption and scattering, and
\revold{$L_{\mu_{\mathrm{a}}}$ and $L_{\mu_{\mathrm{s}}}$}
\revnew{$\mathcal{L}_{\mu_{\rm a}}$ and $\mathcal{L}_{\mu_{\rm s}}$}
are the Cholesky decompositions of the inverse of the prior covariance
matrices for absorption and scattering, 
\revold{$\Gamma_{\mu_{\mathrm{a}}}^{-1}= L_{\mu_{\mathrm{a}}}^{\rm T} L_{\mu_{\mathrm{a}}}$ and $\Gamma _{\mu_{\mathrm{s}}}^{-1}= L_{\mu_{\mathrm{s}}}^{\rm T} L_{\mu_{\mathrm{s}}}$,}
\revnew{$\Gamma_{\mu_{\rm a}}^{-1}= \mathcal{L}_{\mu_{\rm a}}^{\rm T} \mathcal{L}_{\mu_{\rm a}}$ and $\Gamma _{\mu_{\rm s}}^{-1}= \mathcal{L}_{\mu_{\rm s}}^{\rm T} \mathcal{L}_{\mu_{\rm s}}$,}
respectively.

The prior model for the unknown parameters $\muaa$ and $\muas$ was
chosen to be based on the Ornstein-Uhlenbeck process
\cite{rasmussen2006, pulkkinen2014}. 
\revnew{In the authors' previous studies, the prior has been found
  efficient and versatile for multiple type of imaged targets. A
  comparative discussion of the prior selection in applications of
  QPAT is provided in Ref. \cite{pulkkinen2014}.}
\revold{It is}
\revnew{Ornstein-Uhlenbeck prior is} a Gaussian distribution with a covariance matrix $\Gamma$
defined as
\begin{equation}
  \label{eq:prior_covariance}
  \Gamma = \sigma^2 \Xi
\end{equation}
where $\sigma$ is the standard deviation of the prior and $\Xi$ is a
matrix which has its elements defined as
\begin{equation}
  \label{eq:ornstein_uhlenbeck_covariance}
  \Xi_{ij} = \exp( - || r_i - r_j ||  / \ell ),
\end{equation}
where $i$ and $j$ denote the row and column indexes of the matrix,
$r_i$ and $r_j$ denote the coordinates of the discretization points
$i$ and $j$, and $\ell$ is the characteristic length scale of the
prior describing the spatial distance that the parameter is expected
to have (significant) spatial correlation for.

\revnew{The expected value and standard deviation of the prior are chosen such that the
  relevant support of the Gaussian prior describes that of the imaged
  target. These parameters can be chosen, for example, based on
  expected range of variation of the properties of the
  target. Similarly the characteristic length scale $\ell$ is usually
  chosen to be in the same scale as the expected size of heterogeneities
  found in the imaged target. These parameters are application and
  imaged target specific and hence are tunable parameters. Parameter
  choises used in this work are described in Sec.
  \ref{sec:imagereconstruction}.}

Minimization problem \refeqn{eq:MAP_QPAT} can be solved using methods
of numerical optimization.  Here we use the Gauss-Newton method
augmented with a line-search algorithm \cite{Nocedal06}.  A
Gauss-Newton iteration can be written in a form
\begin{equation}
  \begin{split}
    \label{eq:Gauss_Newton}
    x_{(i+1)} = x_{(i)} + s_{(i)}\left(J_{(i)}^{\rm T} \Gamma_e^{-1} J_{(i)} + \Gamma_x^{-1} \right)^{-1} \cdot \quad \qquad \\ \left( J_{(i)}^{\rm T} \Gamma_e^{-1} (H_{\rm meas}-H_{(i)}-\eta_e) - \Gamma_x^{-1} (x_{(i)}-\eta_x)  \right)
  \end{split}
\end{equation}
where 
\revold{$x_{(i)}=(\muaa ,\muas)^{\rm T} = (\mu_{{\rm a}_1},\ldots, \mu_{{\rm a}_K},\mu_{\rm s_{1}},\ldots ,\mu_{\rm s_{K}})^{\rm T} \in \R^{2K}$}
\revnew{$x_{(i)}=(\muaa ,\muas)^{\rm T} = (\mu_{{\rm a}, 1},\ldots, \mu_{{\rm a}, n},\mu_{{\rm s}, 1},\ldots ,\mu_{{\rm s}, n})^{\rm T} \in \R^{2n}$}
are the estimated absorption and scattering parameters and $s_{(i)}$
is the step length at iteration $i$, and
\begin{equation}
  \Gamma_x^{-1} = \begin{pmatrix}
    \Gamma_{\muaa}^{-1} & 0 \\ 0 & \Gamma_{\muas}^{-1} \end{pmatrix},
  \quad \eta_x = \begin{pmatrix}
    \eta_{\muaa} \\ \eta_{\muas} 
  \end{pmatrix}.
\end{equation} 
Further, $J_{(i)}$ is the Jacobian of the form
\begin{equation}
  \label{eq:Jacobian}
  J_{(i)} = \begin{pmatrix}
    J_{\muaa,(i)} & J_{\muas,(i)} 
  \end{pmatrix}
\end{equation}
where $J_{\muaa,(i)}$ and $J_{\muas,(i)}$ are Jacobians for absorption
and scattering.  The formulation of the forward operator and Jacobian
matrices are based on Monte Carlo and PMC methods described in
Sec. \ref{sec:pmcpt}.
%==============================================================
% Section III: Perturbation Monte Carlo method for QPAT
%==============================================================
\section{Perturbation Monte Carlo method for QPAT}
\label{sec:pmcpt}

%===============================================================

\subsection{Monte Carlo method for light transport}
\label{sec:MC}

In Monte Carlo method for light transport in biological tissue, the
underlying model for light propagation has \revold{four}\revnew{three} main principles \cite{prahl89}. 
Firstly, the probability for photon absorption in a small length $\dee s$ in a propagation direction is $\muaa \dee s$. 
Secondly, the probability for photon scattering is similarly $\muas \dee s$. 
Hence, the scattering length follows an exponential
probability distribution function 
\begin{equation}
\label{eq:PDF_scattering_length}
f(l) = \muas(l) \exp \left[- \int_0^l \muas(s) \dee s \right].
\end{equation}
Thirdly, if a scattering occurs, the scattering angle follows a
probability distribution for scattering direction which in this work
is the Henyey-Greenstein phase function
(\ref{eq:henyeygreenstein2Dand3D}). 
\revold{In addition, refractive index changes on the boundaries of the
  domain and sub-domains affect photon propagation and change the
  direction.}

Typically, in order to generate \revold{a good}\revnew{sufficient}
statistics with optimal efficiency, so-called photon packet method is
used \cite{prahl89}.  In the photon packet method, instead of
simulating propagation of individual \revold{photos}\revnew{photons}
until an absorption event, a 'group of photons' (a photon packet) with
an initial weight $w$ is simulated.
As the photon packet propagates, its weight along trajectory $S$\revold{reduces according to absorption contributing continuously}
\revnew{is reduced due to absorption according to}
\begin{equation}
  \label{eq:absorption_weight_reduction}
  w(S) = \exp \left[ - \int_S \muaa (s) \dee s \right]. 
\end{equation}
This is continued until the photon packet exits the simulation domain,
or its weight becomes small.  Sampling scattering lengths from
Eq. (\ref{eq:PDF_scattering_length}) with a weight factor assigned to
those trajectories according to
Eq. (\ref{eq:absorption_weight_reduction}) is a form of importance
sampling \cite{Ivan1991}.  It produces statistically equivalent
results compared to the straightforward generation of photon paths
that can end on absorption events. This method is called the
microscopic Beer-Lambert law in \cite{sassaroli2012}.

In Monte Carlo simulation for QPAT with piecewise constant optical
coefficients $\mu_{{\rm a},i}$ and $\mu_{{\rm s},i}$, the total
absorbed optical energy density $H_{j}$ deposited to discretization
element $j$ is computed as
\revold{$H_{j}=\frac{1}{A_j}\sum_{e \in \{\text{entrances}\}} w_{e} (1-\exp [-\mu_{{\rm a},j} S_{e,j}])$}
\revnew{
  \begin{equation}
    \label{eqn:abs}
    H_{j}=\frac{1}{A_j}\sum_e w_{e}   \left(1-\exp \left[-\mu_{{\rm a},j} S_{e,j}\right]\right)
  \end{equation}
}%
where $w_{e}$ is the weight \revnew{of the photon packet} before
entrance to the $j$:th element, and $S_{e,j}$ is the distance traveled
on each entrance to $j$.  $A_j$ is the area/volume of the element in
2D/3D.  
\revold{Index $e$ refers to each entrance, including revisits, to the
  element as indicated below the sum sign.}
\revnew{The summation over index $e$ refers to each entrance,
  including revisits, by the photon packet to the element.}
\revold{The summation will be hereafter notated simply over $e$ to indicate
each entrance.}

%================================================================

\subsection{Perturbation Monte Carlo}
\label{sec:pmc}

The goal of perturbation Monte Carlo is to evaluate the effect of a small change in the optical parameters (perturbation) to the simulation results efficiently. 
This goal is achieved by re-using the trajectories from an unperturbed simulation so that the trajectories do not need to be re-generated for each perturbation
\cite{Ivan1991}.
Considering the ratio between probability density functions between scattering lengths in perturbed and unperturbed regions of a domain and utilizing the knowledge that scattering length does not depend on absorption, the following expression for the weight of a photon packet in a perturbed simulation\revold{$\tilde{w}$} can be derived
\begin{equation}
  \label{eqn:weight_pert_dicrete}
  \tilde{w} = w \left(\frac{\tilde \mu_{\rm s} }{\mu_{\rm s}} \right )^{k} \exp \left[{-({\tilde \mu}_{\rm s} - \mu_{\rm s}) L_{\rm tot}}\right]
\end{equation}
\revnew{where \revnew{$\tilde{w}$} is the perturbed weight, $w$ is the unperturbed weight,
$\tilde \mu_{\rm s}$ is the perturbed scattering coefficient, }\revold{is the weight of an unperturbed simulation,}$L_{\rm tot}$ is \revold{the trajectory of the photon packet inside of the perturbed region}\revnew{the total distance travelled by the photon packet inside the perturbed region,}
 and $k$ the number of scattering events in the perturbed region.
For more details of the derivation, see Appendix \ref{sec:appx_pmc}.

Now, in QPAT in a perturbed medium with piece-wise constant optical
coefficients $\mu_{{\rm a},i}$ and  $\mu_{{\rm s},i}$, the total energy density $\tilde{H}_{j}$ deposited to discretization element $j$ is 
\begin{equation}
\label{eqn:absorbed_energy_pert_discrete}
  \tilde H_{j} = \frac{1}{A_j} \sum_{{{e}}}\tilde w_{e}\left(w_{e},\tilde{\mu}_{{\rm s},i}, k_{{e},i}, L_{{e},i}\right)  \left(1-\exp\left[-\mu_{{\rm a},j} S_{{e},j}\right]\right)
\end{equation}
where $k_{{e},i}$ is the number of scattering events and $L_{{e},i}$ is 
\revold{the photon trajectory} 
\revnew{the total distance travelled by the photon packet}
in the perturbed element $i$.

%==========================================================================

\subsection{Construction of the Jacobian}
\label{sec:pmcder}

To construct the Jacobian for the Gauss-Newton algorithm (\ref{eq:Gauss_Newton}), derivatives of absorbed optical energy with respect to the optical coefficients need to be evaluated. 
The derivative for the absorption coefficient can be computed directly from 
Eq. \refeqn{eqn:abs} by differentation.
For construction of the Jacobian for the scattering, perturbation Monte Carlo is utilized, and 
the derivative with respect to the scattering coefficient is computed using
Eqs. (\ref{eqn:weight_pert_dicrete}) and (\ref{eqn:absorbed_energy_pert_discrete}).

Here, in the case of piece-wise constant absorbed optical energy density $H_j$ and optical parameters $\mu_{{\rm a},i}$ and $\mu_{{\rm s},i}$, the derivatives can be derived to take the following forms.
For absorption, if $i \neq j$,
\begin{equation}
  \label{eqn:nondiagonalmua}
  \left (\frac{\partial H_{j}}{\partial \mu_{{\rm a},i}} \right )_{i \neq j} = \frac{1}{A_j} \sum_{e} -L_{{e},i} w_{e} \left(1-\exp\left[-\mu_{{\rm a},j} S_{{e},i}\right]\right)
\end{equation}
and, if $i = j$,
\begin{equation}
\begin{split}
  \label{eqn:diagonalmua}
  \frac{\partial H_{j}}{\partial \mu_{{\rm a},j}} = & \frac{1}{A_j} \sum_{e} w_{e} \left( L_{e,i} \left(\exp\left[- \mu_{{\rm a},j} S_{{e},j}\right] - 1\right) \right. \\ & \left. +  S_{{e},j} \exp \left[-\mu_{{\rm a},i} S_{{e},j}\right]\right).
\end{split}
\end{equation}
For scattering 
\begin{equation}
\begin{split}
  \label{eqn:musmua}
  \frac{\partial H_j}{\partial {\mu_{{\rm s},i}}}  = & \frac{1}{A_j} \sum_{e} w_{e} \left(\frac{k_{{e},i}}{\mu_{{\rm s},i}} - L_{{e},i} \right) \\ & \cdot \left(1 -\exp\left[-\mu_{{\rm a},j} S_{{e},j}\right]\right).
\end{split}
\end{equation}
Derivation of absorption and scattering derivatives and their computational validation are presented in Appendixes \ref{sec:appx_jacobian_construction} and  \ref{sec:appx_validation_jacobian}.

%========================================================================
% Section IV: Simulations
%========================================================================

\section{Simulations}
\label{sec:simulations}

The perturbation Monte Carlo approach for QPAT was evaluated with
numerical simulations.  Two types of simulations were
considered. First, only the optical problem was studied to evaluate
the performance of the proposed method with different targets and
noise levels, and then, the full photoacoustic simulation including
also the acoustic part was performed\revnew{.}
\revold{to simulate a more realistic imaging situation.}
\revnew{While idealized, the purpose of the latter simulation is to
mimick a more realistic imaging scenario by a fine-detailed,
biomedical target and a finite number of acoustic sensors.}
In all cases, the simulation domains were two-dimensional squares with
edge length of $5 \, {\rm mm}$.
\revnew{Multiple different illuminations were used to ensure that the
  optical absorption and scattering can be estimated. The imaged
  target was illuminated by four different illumination patterns
  originating from the four sides of the domain respectively.}

In the first simulation concentrating only on the optical inverse problem of QPAT, two targets were studied.
The absorption and scattering parameters of these targets are shown in Figs. \reffig{fig:reco1} and \reffig{fig:reco2}.
The first target, i.e. 'bars' (Fig. \reffig{fig:reco1}), featured 4 bars with absorption values $0.05, \, 0.02, \, 0.005$
and  $0.0001 \, {\rm mm}^{-1}$ and  scattering  values $0.01, \, 0.5, \, 2$ and $5 \, {\rm mm}^{-1}$. 
The background absorption coefficient was $0.01 \, {\rm mm}^{-1}$ and the background scattering coefficient was $1 \ {\rm mm}^{-1}$. 
In the second target, i.e. 'cards' (Fig. \reffig{fig:reco2}), the absorption coefficients of the clover and spade were $0.05$ and $0.02 \, {\rm mm}^{-1}$, respectively, and the scattering coefficients of the spade and heart were $2 \, {\rm mm}^{-1}$.  The diamond had a spatially varying
 scattering coefficient between 0.8 and 1.8 ${\rm mm}^{-1}$.
The background absorption coefficient was $0.01 \, {\rm mm}^{-1}$ and the background scattering coefficient was $1 \, {\rm mm}^{-1}$.
In all simulations, the anisotropy parameter of the \revold{Henyey-Greenstain}\revnew{Henyey-Greenstein} phase function was $g=0.9$.
The refractive index of the target was a constant and matched with the refractive index of surrounding medium, i.e. no light reflections on the boundary or within the target occurred.

In the full photoacoustic simulation, a blood-vessel mimicking numerical \revold{fantom}\revnew{phantom} of the k-Wave toolbox illustrated in Fig. \ref{fig:reco_vessel} was studied. 
In that case, the absorption coefficients of the vessel and background were $0.4$  and $0.07  \, {\rm mm}^{-1}$, respectively, and the scattering coefficients of the vessel and the background were  $20$ and $9 \, {\rm mm}^{-1}$, respectively.
These values correspond approximately to absorption and scattering of blood and adipose tissue at wavelenghth $\lambda \approx 900 \, {\rm nm}$ \cite{jacques2013,friebel2006}. 
The anisotropy parameter of the \revold{Henyey-Greenstain}\revnew{Henyey-Greenstein} phase function was again $g=0.9$,  and the refractive index of the target was a constant matching with the refractive index of the surrounding medium. 

In addition to visual inspection of the simulation results, we compared  the differences between estimated and ground truth values using 
\begin{equation}
  \label{eqn:error}
  E = 100 \%  \cdot \sqrt{ \frac{  \int_{\Omega} ( f(r) - f_{\text{ref}}(r) )^2 \dee  r } { \int_{\Omega} f_{\text{ref}}(r)^2 \dee r} }
\end{equation}
where $f$ is a discrete presentation of the parameters being estimated at point $r$, i.e. absorption or scattering
coefficient, and $f_{\rm ref}$ is the reference (ground truth value) mapped to the
reconstruction mesh for comparison.

%=============================================================

\subsection{Data generation}
\label{sec:datagen}

The simulation domains were discretized into triangular elements using
NetGen \cite{schoberl1997netgen} software for the 'bars' and 'cards'
simulations.\revold{In the case of the 'vessel' simulation, the $350 \times 350$ pixel
grid of the original image was split into two identical triangles.}
\revnew{The 'vessel' simulation used a $350 \times 350$ pixel grid, where the pixels were spit into two identical triangles.}
The number of elements and nodes in each mesh is
given in Table \reftable{table:forward_mesh}.

\begin{table}[!tp]
\center
  \caption{Number of elements and nodes of the discretization, and the number of photon packets used in each
  illumination in data simulation.}
  \begin{tabular}{lrrc} \hline 
  Target  & Elements & Nodes & Packets  \\
  \hline
  'bars' (Fig. \reffig{fig:reco1})  & $92182$ & $46494$ &$10^9$   \\
  'cards' (Fig. \reffig{fig:reco2}) & $89516$ & $45161$ &$10^9$   \\
  'vessel' (Fig. \reffig{fig:reco_vessel}) & $245000$ & $123201$ & $10^8$ 
  \\ \hline
  \label{table:forward_mesh}
  \end{tabular} 
  
\end{table}

Photon fluence and absorbed optical energy density were simulated with the Monte Carlo method implemented with ValoMC MATLAB toolbox \cite{Leino19}.
The targets were illuminated from the left, right, bottom and top faces
separately with a collimated light source (i.e. all photon packets travel initially in the direction of the face normal \revnew{and the whole face acted as a light source}). 
The number of photon packets used in simulations for each illumination is given in Table \ref{table:forward_mesh}.

In the pure optical simulation with 'bars' and 'cards', the absorbed optical energy density within the domain was stored as data. 
Two separate noisy datasets (for both targets) were obtained by adding uncorrelated Gaussian noise with
standard deviation equal to $1 \, \% $ or $0.1 \, \% $  of the maximum value of the simulated data.
In addition, Monte Carlo simulations contain also intrinsic noise. 
To estimate the intrinsic noise level for these targets, we ran an additional forward simulation and computed the standard deviation of the difference between the two results. 
These values were small compared to the
added noise (less than $0.03 \, \%$ of the maximum for both targets). 

In the full photoacoustic simulation with the 'vessel' target, we
continued from the absorbed energy density obtained from the optical
simulation by computing the initial pressure using
Eq. (\ref{eq:initial_pressure})\revnew{.}\revold{with Gr{\"u}neisen parameter $G = 0.1$} 
\revnew{In this work we assumed a known Gr{\"u}neisen parameter. This
  makes mapping between the initial pressure and the optical absorbed
  energy density trivial, and thus the choice of its value arbitrary.}
The initial pressure was computed in \revold{the} $1050 \times 1050$
pixel grid, that was then used to simulate the photoacoustic wave
propagation.  The acoustic initial value problem
(\ref{eq:wave_equation_pat}) was solved using the $k$-space
time-domain method implemented with the k-Wave toolbox
\cite{treeby2010a}.  For the speed of sound, value $c = 1500 \, {\rm
  m/s}$ was used, which is similar to the speed of sound in water and
soft tissues.  The time-varying acoustic pressure was recorded at\revold{4196}\revold{sensors located at the boundary of the grid.}
\revnew{600 sensors spread uniformly on the boundary of the grid with
  sensors being separated by $33 \: \mathrm{\mu m}$. Ideal point like
  sensors were used.}
A perfectly matched layer \cite{treeby2010a} with 
\revold{350}
\revnew{50} 
grid points in each dimension was used outside of grid to dampen the
escaping waves. The pressure signals were sampled for\revold{$12 \times 10^{-6}$ seconds}
\revnew{$12 \: \mathrm{\mu s}$}
and discretized into $12602$ temporal points at each
of the acoustic sensor locations. 
% \revnew{Ideal transducers point-like were used assumed. The spectra of the signals contained weak peaks up to 150 MHz.}
Noise with a standard deviation of $1 \, \% $ of the peak amplitude of
the simulated pressure signal was added to the simulated data.  This
type of sensor positioning and noise can roughly be regarded to
simulate a Fabry-P{\'e}rot based photoacoustic sensor-head
\revnew{\cite{zhang2008a,buchmann2016,ellwood2017}.}
In order to obtain the data for the optical inverse problem, the
acoustic inverse problem was solved using the time-reversal method
implemented with the k-Wave toolbox.  Discretization of $850 \times
850$ pixels was used.  As a result, the initial pressure within the
domain was obtained.\revold{Then, the absorbed optical energy density was computed from
  the initial pressure using Eq. (\ref{eq:initial_pressure}) with the
  Gr{\"u}neisen parameter $G=0.1$, }
\revnew{Then, the absorbed optical energy density can be obtained
  trivially from the initial pressure using the known Gr{\"u}neisen
  parameter,}
and regridded to the discretization of the optical inverse problem.

%===========================================================

\subsection{Image reconstruction}
\label{sec:imagereconstruction}

In the solution of the inverse problem the photon fluence and absorbed
optical energy density were represented in a mesh constructed of
regular triangular elements (two triangles form a rectangle) with $20
000$ elements and $10 201$ nodes.  The absorption and scattering
parameters were represented in a $100 \times 100$ rectangular pixel
grid.  It should be noted that, even though we used regular meshes in
the solutions of the inverse problem, it could be solved in an
irregular mesh as well.

\revold{The MAP estimates \refeqn{eq:MAP_QPAT} of the optical parameters were
computed.} In the solution of the inverse problem, the measurement
noise statistics were assumed to be known. For the 'bars' and the
'cards', simulated $1 \, \% $ or $0.1 \, \% $ noise levels were used,
while for the 'vessel' a $1 \, \% $ noise 
\revold{is} 
\revnew{was}
used.  For the
Ornstein-Uhlenbeck prior, the expected values $\eta_{\muaa}$ and
$\eta_{\muas}$ were\revold{set as the background optical parameters for'bars' and 'cards', whereas for the 'vessel', they were}
set to the midpoint value between the maximum and the minimums.
The standard deviations $\sigma_{\muaa}$ and $\sigma_{\muas}$ were set
such that maximum target values corresponded to two standard
deviations from the prior mean.  The characteristic length scale was
set as $0.5 \, {\rm mm}$ for the 'bars' and 'cards' targets and as
$0.1 \, {\rm mm}$ for the 'vessel', corresponding approximately to the
size of the inhomogeneities.  

The MAP estimates were computed using
the Gauss-Newton method.  As the initial guess for absorption and
scattering in the Gauss-Newton iteration (\ref{eq:Gauss_Newton}), the
expected values of the prior were used.  During the iterations, the
forward solutions and the Jacobians were computed using Monte Carlo
and PMC as described in Sec. \ref{sec:pmcpt}.  
\revnew{To optimize the computational efficiency, the forward model
  and the Jacobian were evaluated in a single simulation utilizing the
  same photon packet trajectories.}
The number of photon packets used to produce the forward solution
and Jacobian
\revnew{during the Gauss-Newton iterations}
was $10^8$ for each illumination of the 'bars' and
'cards' and $10^7$ for the 'vessel'.
 
Difference between consecutive reconstructions was computed using
Eq. \refeqn{eqn:error}, where $f$ were the estimated values and
$f_{\text{ref}}$ were the estimates of a previous iteration.  The
Gauss-Newton iteration was stopped once the average difference between
three consecutive reconstructions \revold{the estimates of the
  previous iteration}had fallen below $0.5 \, \%$.

%===================================================================

\subsection{Results}
\label{sec:results} 

\begin{figure}[tp!]
  \begin{center}
  %\begin{tabular}{c}
  \includegraphics[width=8.5cm]{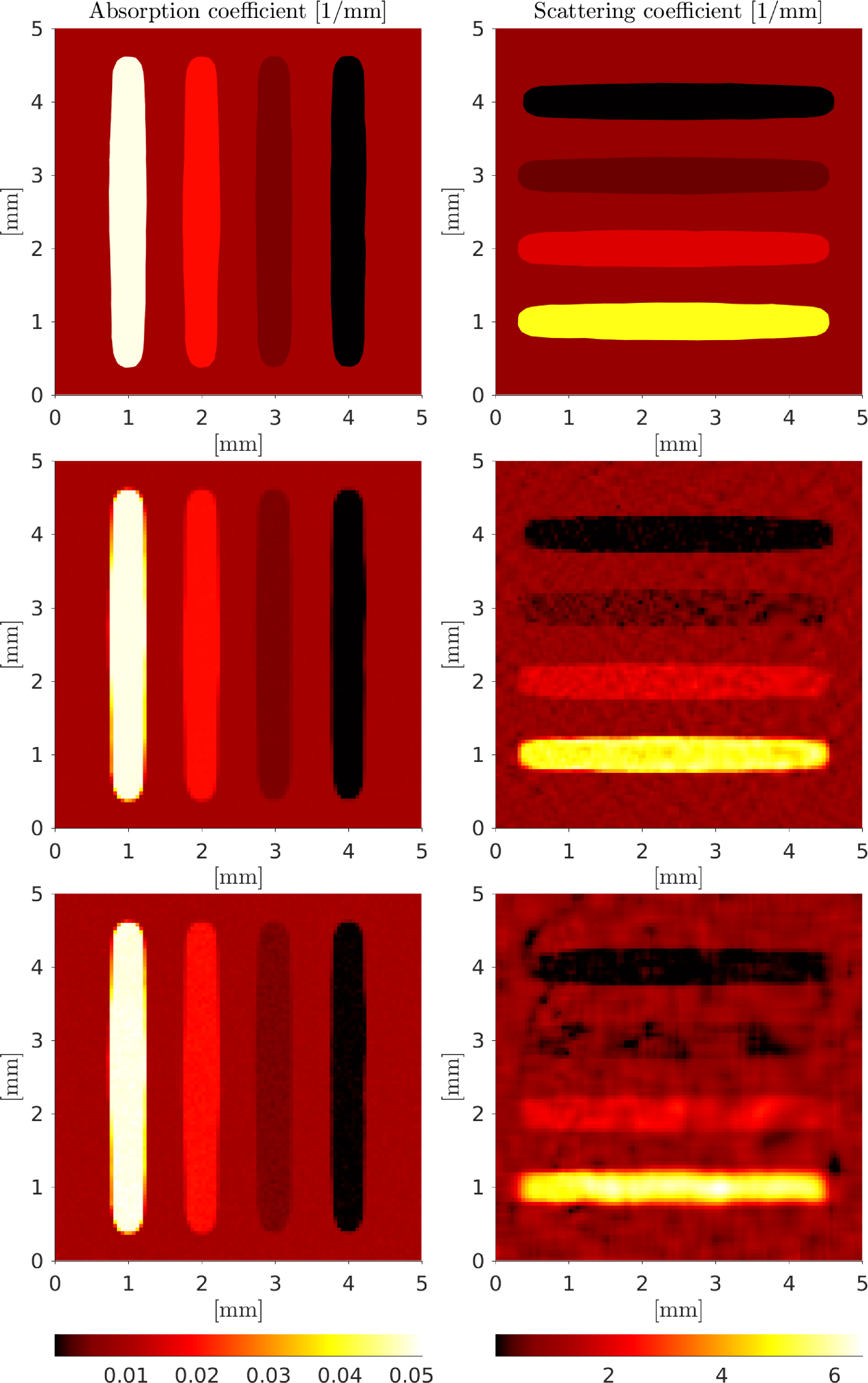}
  %\end{tabular}
  \end{center}
  \caption 
  {\revnew{Reconstructed absorption (left column) and scattering (right column) distributions. Rows from top to bottom:  simulated true values (first row), reconstructions from data with $0.1 \, \%$ of additive noise (second row) and reconstructions from data with $1 \, \%$ of additive noise (third row).}} 
\label{fig:reco1}
\end{figure}

\begin{figure}[tp!]
  \begin{center}
  %\begin{tabular}{c}
  \includegraphics[width=8.5cm]{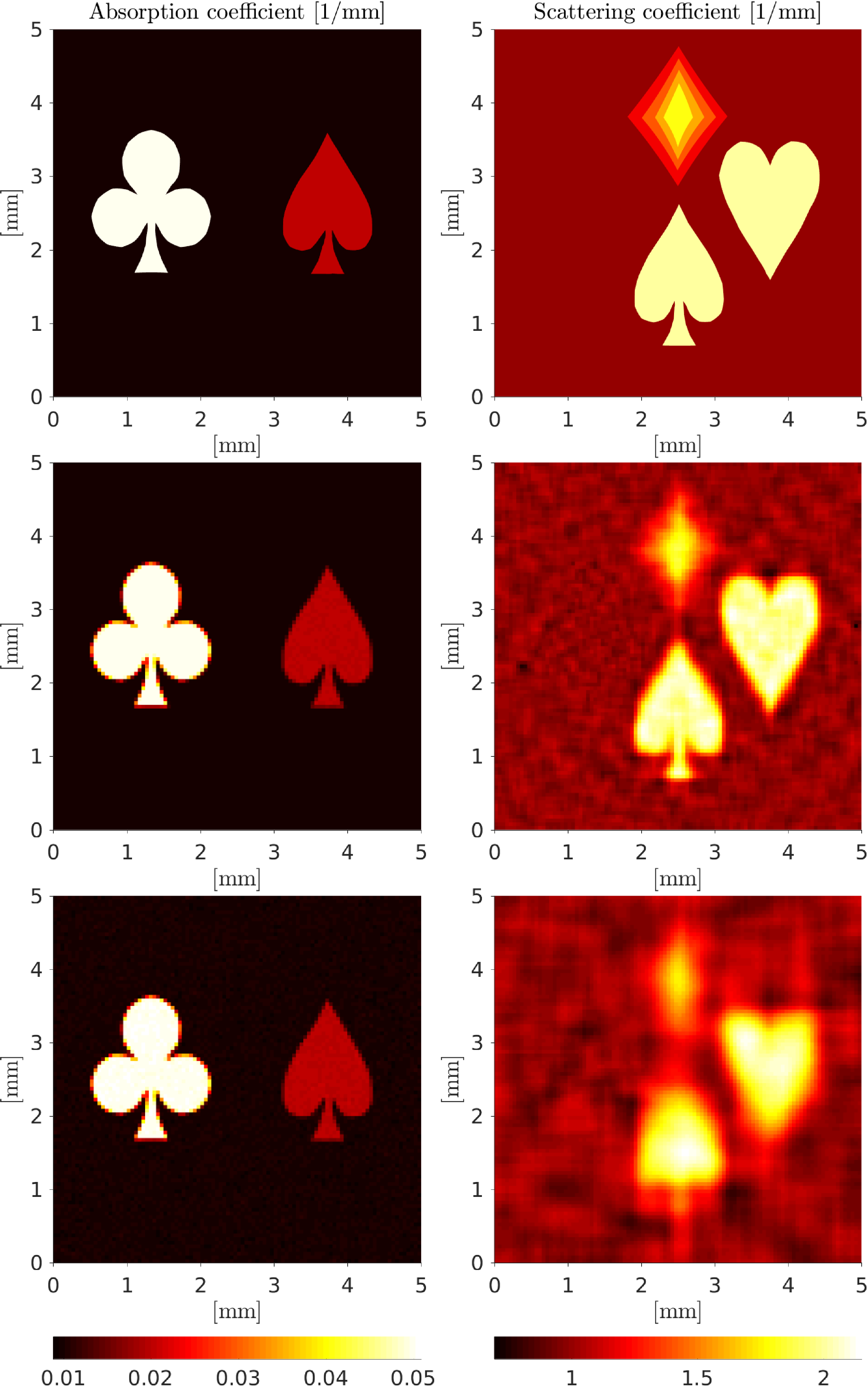}
  %\end{tabular}
  \end{center}
  \caption 
  {\revnew{Reconstructed absorption (left column) and scattering (right column) distributions. Rows from top to bottom:  simulated true values (first row), reconstructions from data with $0.1 \, \%$ of additive noise (second row) and reconstructions from data with $1 \, \%$ of additive noise (third row).} }
  \label{fig:reco2}
\end{figure}

The reconstructed absorption and scattering coefficients for 'bars'
and 'cards' targets with both noise levels are shown in
Figs. \reffig{fig:reco1} and \reffig{fig:reco2}.  As can be seen, in
all cases the absorption coefficients are well reconstructed and
represent the target features well.  Based on qualitative inspection,
the noise level does not have a significant impact on the absorption
estimates.  In the reconstructions on scattering, however, the impact
of noise is clearly visible, although also in these the main features
of the target are quite well captured.  This is especially evident in
the cards experiment shown in Fig. \reffig{fig:reco2}.  Some artefacts
are seen in particular in the region with the lowest absorption
coefficient in Fig. \reffig{fig:reco1}.  

We refrain from making
definite judgments about the performance in relation to other
reconstruction methods due to differences in the simulations.
However, compared to the previous
\cite{tarvainen2012,saratoon2013,hochuli2016} reconstructions obtained
using the RTE or Monte Carlo as the forward model, those presented
here do seem to show a promising
\revold{quality} 
\revnew{quantitative accuracy} in general. 
\revnew{Results that enable a more straightforward comparison with the
  closely related adjoint Monte Carlo method\cite{hochuli2016} are
  presented in the supplementary material. The comparison favors PMC,
  but it should not be taken as a proof of superiority. This is
  because computation expense can vary greatly between the methods, which
  is discussed in the supplementary material.}

\begin{figure}[tp!]
  \begin{center}
  \begin{tabular}{c}
  \includegraphics[width=8.5cm]{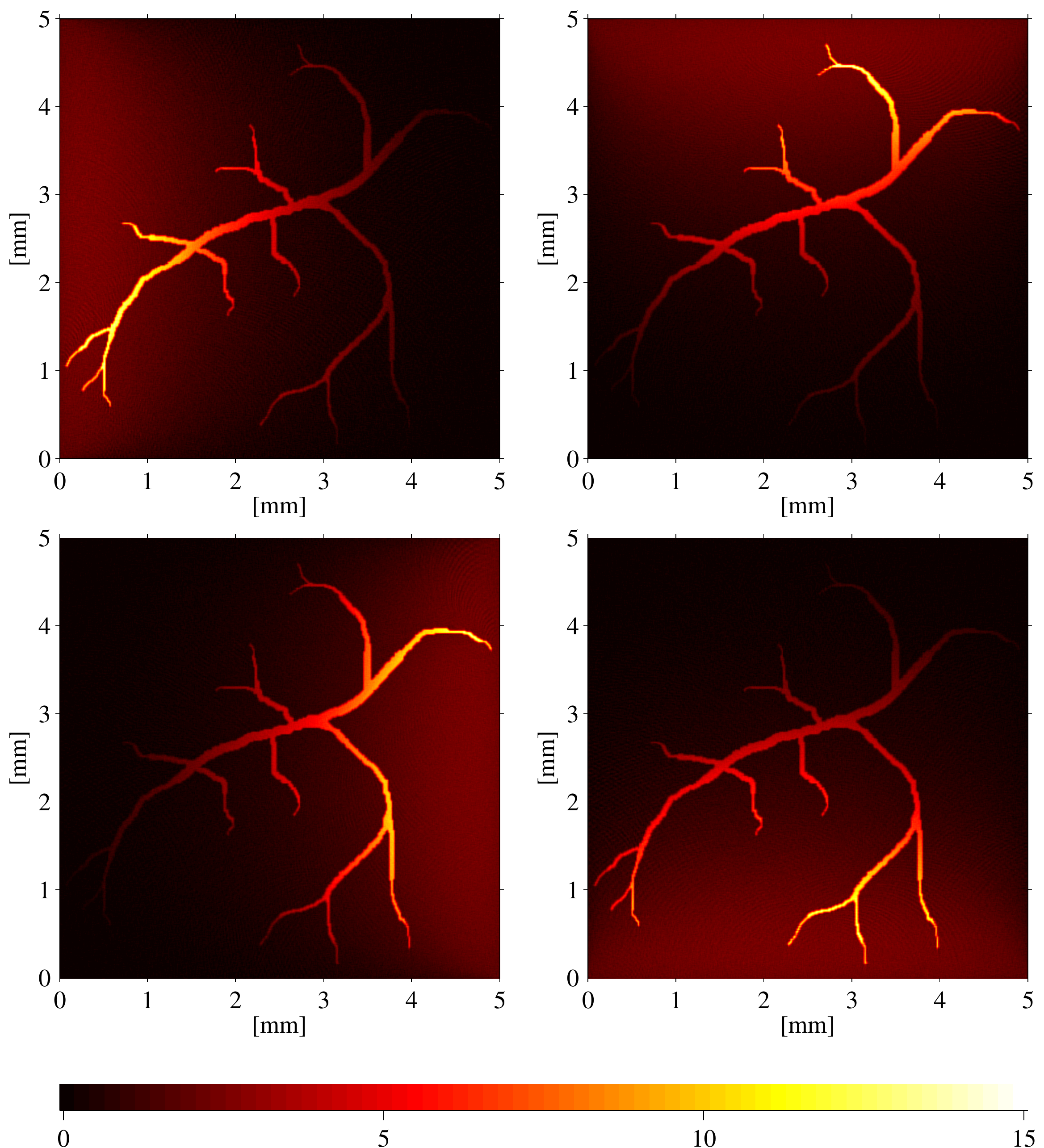}
  \end{tabular}
  \end{center}
  \caption 
  {\revnew{Reconstructed initial pressure distributions (arbitrary units) for the four
      illuminations of the 'vessel' target. Reconstructions shown for
      illumination originating from the left (top left), top (top right), right
      (bottom left), and bottom (bottom right).}}
  \label{fig:pressures_vessel}
\end{figure}

\revnew{For the 'vessel' target, the reconstructed initial pressure distributions are shown in
  Fig. \ref{fig:pressures_vessel}. The estimates resemble the true
  distributions (data not shown) qualitatively with a smooth absorbed
  energy density field emanating from the light source overlaid with
  strongly absorbing heterogeneity mimicking blood-vessels. The
  relative errors of the estimated initial pressure distributions were
  in range $7.7-9.9\%$.}

\begin{figure}[tp!]
  \begin{center}
  \begin{tabular}{c}
  \includegraphics[width=9cm]{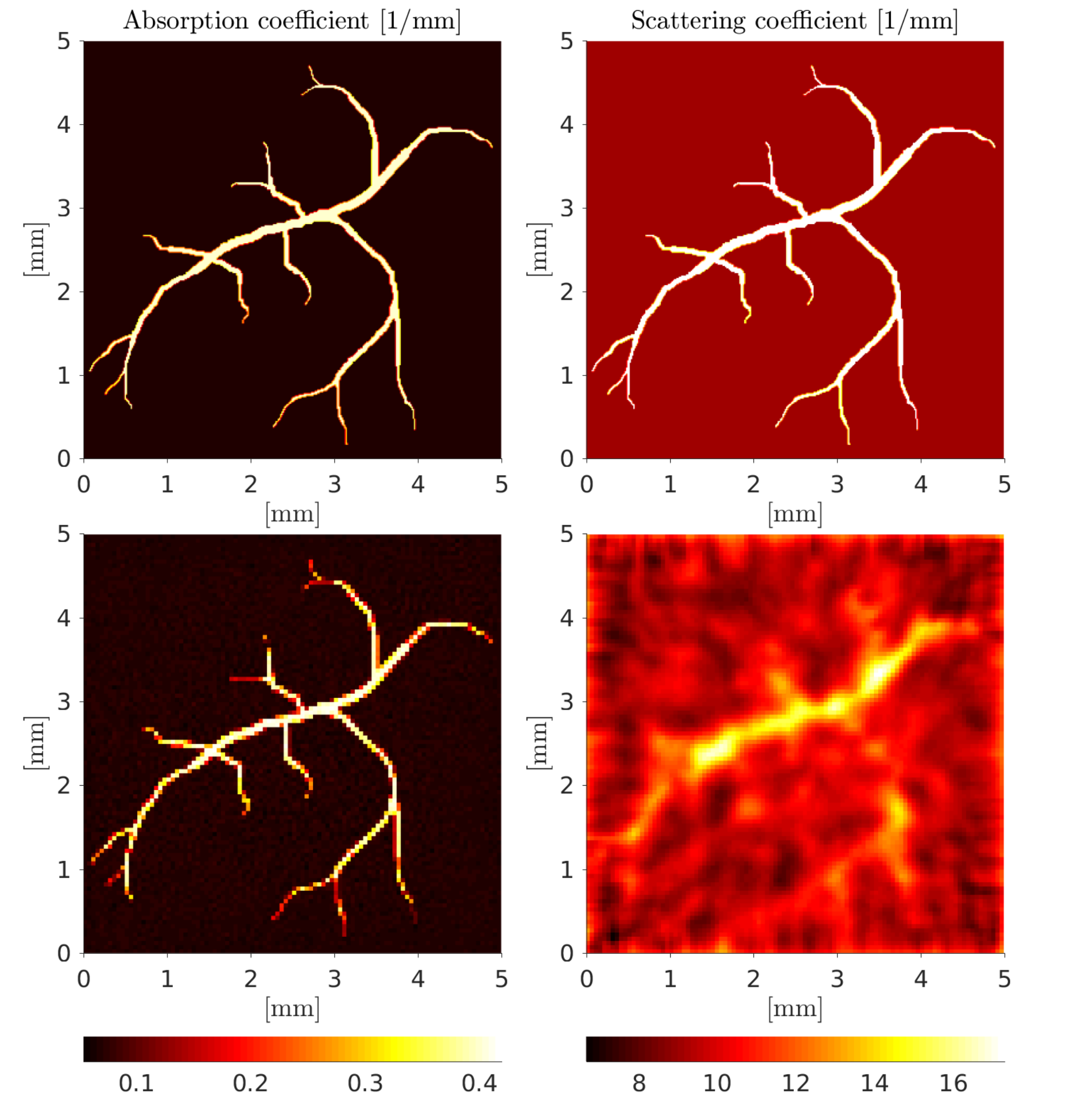}
  \end{tabular}
  \end{center}
  \caption 
  {\revnew{Reconstructed absorption (left column) and scattering (right column) distributions. Simulated true values (top row) and  reconstructions from data with $1 \, \%$ of additive noise (bottom row).}} 
  \label{fig:reco_vessel}
\end{figure}

\revold{The} 
\revnew{Further, the 
reconstructed absorption and scattering coefficients for the
'vessel' target are shown in Fig. \ref{fig:reco_vessel}.}
As it can be seen, the main features of the target are captured well
both in the absorption and scattering images.  The absorption image
represents clearly also the small vessels.  The scattering image, on
the other hand, is not as sharp as the absorption, which was noticed
also in other simulations.

The relative errors of the estimates computed using
Eq. \refeqn{eqn:error} are given in Table \reftable{table:error}.  As
can be seen, the quantitative accuracy of the absorption estimates is
good for all tests.  The relative errors are approximately on the same
level for both 'bars' and 'cards' tests with the same amount of
additive noise with lowest error of $0.2 \, \%$ obtained with 'cards'
test with $0.1 \, \%$ of noise and highest error of $2.2 \, \%$
obtained with 'bars' test with $1 \, \%$ of noise.
The relative errors of scattering are larger, and only in the 'cards'
test with lower amount of additive noise the error is below $10 \,
\%$.  In the case of the 'vessel' simulation, the relative errors of
both absorption and scattering are slightly larger than in the other
simulations.
%
%% Esimerkki tilanteesta jossa revold on huono, tama piippaa virhetta. Merkattu eritavalla
%
\revold{This is explained by the numerics of the acoustic simulation
  that affects both data and noise. Similar effect was also noticed
  in}
%
%{\color{red}\cite{tarvainen2013}}
%
\revold{ where modeling of noise and errors of the acoustic solver was
  studied.}

\begin{table}[!tp]
\center
  \caption{Number of iterations needed to reach convergence of the reconstruction algorithm and the relative errors of the absorption $E_{\muaa} (\%)$ and scattering  $E_{\muas} (\%)$ estimates evaluated using Eq. \refeqn{eqn:error} in different simulations.} \label{table:error}
  \begin{tabular}{lcccc} \hline 
  Target      & Noise level (\%)& Iterations &  $E_{\muaa} (\%)$ & $E_{\muas} (\%)$ \\
  \hline
  'bars' (Fig. \reffig{fig:reco1})  & 0.1            & \revold{19}\revnew{18}                                   & 0.3        & \revold{15}\revnew{11}       \\
  'bars' (Fig. \reffig{fig:reco1})  & 1.0             & \revold{12}\revnew{10}                                   & 2.2         & \revold{25}\revnew{20}          \\
  'cards' (Fig. \reffig{fig:reco2}) & 0.1             & \revold{10}\revnew{9}                                   & 0.2         & \revold{7.0}\revnew{6.1}        \\
  'cards' (Fig. \reffig{fig:reco2}) & 1.0            & \revold{10}\revnew{7}                                   & \revold{1.8}\revnew{1.9}         & \revold{12}\revnew{11}    \\
  'vessel' (Fig. \reffig{fig:reco_vessel}) & 1.0  & \revold{7} \revnew{10} & \revold{2.9} \revnew{5.5}  & \revold{19} \revnew{20}  \\  \hline \end{tabular} 
\end{table}

The computations were performed on a computer with two Intel(R)
Xeon(R) Gold 6136 processors with a total of $24$ cores and $256 \,
{\rm GB}$ of system memory.  The reconstructions took about \revold{$130$} \revnew{$150$}
minutes per step, out of which \revold{$5$--$7$} \revnew{$20$--$30$} minutes \revold{is} \revnew{was} spent outside of
PMC computation (e.g. in the line search and matrix inversions) for
the 'bars' and 'cards'.  For the 'vessel' simulation, the
reconstruction times were longer due to the higher scattering of the
target which resulted in longer Monte Carlo simulation times.  This
time consuming nature was eased by reducing the number of photon
packets which still remained large enough to maintain the numerical
accuracy of the method.  Since it took from 10 to 30 steps to reach
convergence (see Table \reftable{table:error}) and larger
discretization is needed in a realistic geometry, the computation
times are long for practical applications.  However, the photon counts
were based on a simple criterion and not optimized.  Moreover, the
method is trivially parallelizable and easy to implement.  Therefore,
it seems plausible that implementations using e.g. GPU and adaptive
meshing could well succeed in reducing the computation time by orders
of magnitude.\revold{while maintaining the quality.} A systematic
study of the \revold{quality} \revnew{quantitative accuracy} of the
reconstructions as a function of photon packet count remains future
work.

%==============================================================
% Section V: Conclusion 
%===============================================================

\section{Conclusion}
\label{sec:conclusions}    

In this work, a perturbation Monte Carlo method was introduced
  for the optical inverse problem of quantitative photoacoustic
  tomography. Bayesian framework was used for the formulation of the
  inverse problem resulting in a minimization problem that was solved
  using the Gauss-Newton method. 
\revnew{ 
 In the proposed approach,
  Monte Carlo was used as a forward model and perturbation Monte Carlo
  was used to evaluate the Jacobian matrices for the absorbed optical energy
  density with respect to the optical absorption and
  scattering. Evaluation of both the forward model and the Jacobians can
  be implemented in a simultaneous fashion. Although Bayesian framework  was utilised in this work, the perturbation Monte Carlo is not limited to that.  It can be
  utilized similarly with regularization approaches and with different prior models.}
\revold{However, PMC should also work well with other inverse methods
  and minimization algorithms.

The PMC approach was evaluated with numerical simulations.  It was
shown that spatial distributions of both absorption and scattering can
be estimated simultaneously with good accuracy.  One of the main
benefits of the PMC approach is the simplicity.  The part that
computes the derivatives can be implemented in just a few lines of
computer code.}
\revold{The future work includes in addition to optimization of the
method its extension to three dimensions.}
\revnew{Future work includes extending the method to three dimensions.
  In addition, evaluating efficient ways to implement perturbation
  Monte Carlo using graphics processing units (GPU) should be sought
  as the approach is (at least in principle) trivially
  parallelizable. However, potential pitfalls with the GPUs can
  include inefficient memory usage or unideal parallelization
  efficiency.}
Furthermore, \revold{for practical applications,} the method should be
evaluated with experimental data.

\begin{appendices}

  \section{Perturbation Monte Carlo}
  \label{sec:appx_pmc}
 
  To simulate energy absorption $E$ into a domain (or in general to approximate the solution of the RTE \cite{Ivan1991}), one integrates a function $g$ of a trajectory $s$ over probability distribution function (PDF) of  $f$
  \begin{equation}
  \label{eq:appx_energy_PDF}
  E = \int_{S} g(s)f(s)\dee s
  \end{equation}
  where $g(s)$ describes energy absorption and $S$ the 'space' of all photon trajectories. 
  According to the law of large numbers,
  \begin{equation}
  \label{eq:appx_LLN}
  \sum_{n = 1}^N g(s_{n}) \to E, \quad {\rm when} \quad N  \to \infty 
  \end{equation}
  and $s_{n}$ are samples drawn from $f$.
  
  In case we would want to use another PDF $\tilde{f}$ instead, we can obtain these by 'correcting' samples drawn from $f$ by weighting, and write Eq. (\ref{eq:appx_energy_PDF}) as
  \begin{equation}
  \label{eq:appx_energy_PDF_modified}
  E = \int_{S} g(s) \tilde{f}(s) \dee s = \int_{S} g(s) \frac{\tilde{f}(s)}{f(s)} f(s)\dee s
  \end{equation}
  and similarly as in the approximation (\ref{eq:appx_LLN}) we can write
  \begin{equation}
  \label{eq:appx_LLN_modified}
  \sum_{n = 1}^N g(s_{n}) \frac{\tilde{f}(s_{n})}{f(s_{n})} \to E, \quad {\rm when} \quad N  \to \infty 
  \end{equation}
  where $s_{n}$ are samples drawn from $f$.
  
\begin{figure}[!tp]
    \begin{center}
    \includegraphics[width=6cm]{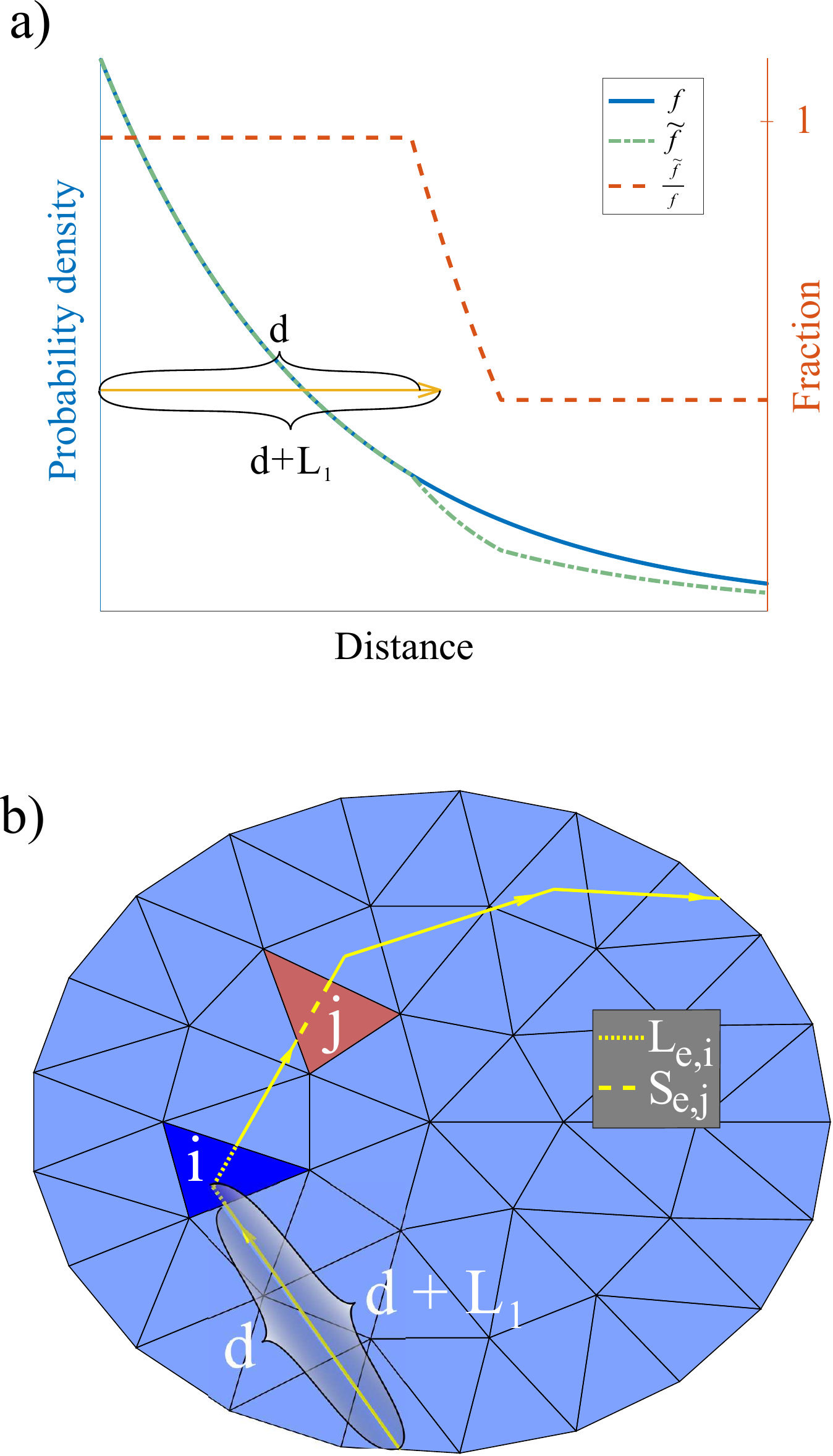}
    \end{center}
    \caption{
      a) Example PDFs for determining the first scattering length
         in the original ($f$) and perturbed simulation ($\tilde f$) for a domain with a piece-wise constant scattering coefficient. 
         Shown is also their ratio that determines the correction factor. 
      b) Domain geometry for the PDFs. Shown is also an exemplar path of a packet. Note that the PDFs are only valid for the first line
       in the path. The index of the perturbed region is $i$ and the region in which the energy density is evaluated $j$.
       The length traversed in $i$ before entrance to $j$ is given
       by $L_{e,i}$, and the length traversed in $j$ by
       $S_{e,j}$. The number of scattering events in the perturbed region is 1 ($k_{e,i} = 1$).
  } 
      \label{fig:pmcschema}
  \end{figure}

  Considering a photon trajectory consisting of multiple steps with a
  scattering distance $L_i$, a modified weight $\tilde{w}$ utilizing
  photon trajectories in perturbed and unperturbed regions can be
  derived
  \begin{equation}
    \label{eqn:appx_wmus2}
    \tilde{w} = w \left(\frac{\tilde{\mu}_{\rm s} }{\mu_{\rm s}} \right )^{k} \exp\left[-\left({\tilde \mu}_{\rm s} - \mu_{\rm s}\right) L_{\rm tot}\right]
  \end{equation}
  where $w$ is the weight of an unperturbed simulation, $L_{\rm tot}$ is
  the total trajectory inside the perturbed region and $k$ the number of
  scattering events in the perturbed region.  The correction factor to
  the weight is the ratio of the PDFs as indicated by
  Eq. (\ref{eq:appx_LLN_modified}).  An example for a piece-wise domain
  is outlined in Fig. \ref{fig:pmcschema}.  
  \revnew{Eq.
    (\ref{eqn:appx_wmus2}) is valid for any range of
    perturbation. However, in optically challenging simulations, the variance of the Monte Carlo simulation can increase
    and an increased number of photon packets may be required to reach
    an acceptable level of noise.}
  
  Now, in QPAT in a perturbed medium with piece-wise constant optical
  coefficients $\mu_{{\rm a},i}$ and $\mu_{{\rm s},i}$, the total
  optical energy density $\tilde{H}_{j}$ deposited to discretization
  element $j$ is (c.f. Eq. \refeqn{eqn:abs})
  \begin{equation}
  \label{eqn:appx_abspert}
    \tilde{H}_{j} = \frac{1}{A_j} \sum_{{{e}}}\tilde{w}_{e}(w_{e},\tilde{\mu}_{{\rm s},i}, k_{{e},i}, L_{{e},i})  \left(1-\exp\left[-\mu_{{\rm a},j} S_{{e},j}\right]\right)
  \end{equation}
  where $k_{{e},i}$ is the number of scattering events and $L_{{e},i}$
  is the photon trajectory in the perturbed element $i$ (see
  Fig. \ref{fig:pmcschema}).  Further, $S_{e,j}$ is the distance
  traveled on each entrance in element $j$ and $A_j$ is the area/volume
  of the element in 2D/3D.  
  %
  %% Index $e$ refers to each entrance\revold{, including revisits,} to the
  %% element as indicated below the sum sign. \revnew{Note these include
  %%   revisits as well}.
  %
  \revold{Index $e$ refers to each entrance, including revisits, to the
    element as indicated below the sum sign.}
  \revnew{The summation over index $e$ refers to each entrance,
    including revisits, by the photon packet to the element.}
  
  %==============================================================
  
  \section{Construction of Jacobian}
  \label{sec:appx_jacobian_construction}

  The derivative for the absorption coefficient can be computed directly
  from Eq. \refeqn{eqn:abs} by differentation, but all occurences of
  $\mu_{\rm a,i}$ must be made explicit since they are only implicitly
  affecting $w_{e}$.  This results in
  \begin{equation}
    \label{eqn:appx_explicitabs}
    H_{j}=\frac{1}{A_j} \sum_{{{e}}} w_{{e}'} \exp\left[-\mu_{\rm a,i} L_{{e},i}\right]  \left(1-\exp\left[-\mu_{\rm a,j} S_{{e},j}\right]\right)
  \end{equation}
  where $w_{{e}'}$ is the weight before entrance to the $j$:th element
  without the contribution from the $i$:th element, and $L_{{e},i}$ is
  the distance traveled in element $i$ before each entrance.  For $i
  \neq j$, differentation gives
  \begin{equation}
    \label{eqn:appx_nondiagonalmua}
    \left (\frac{\partial H_{j}}{\partial \mu_{{\rm a},i}} \right )_{i \neq j} = \frac{1}{A_j} \sum_{e} -L_{{e},i} w_{e} \left(1-\exp\left[-\mu_{{\rm a},j} S_{{e},i}\right]\right)
  \end{equation}
  where $w_e$ contains the contribution that was excluded from $w_{{e}'}$ in Eq. \refeqn{eqn:appx_explicitabs}. If $i = j$,
  \begin{equation}
  \begin{split}
    \label{eqn:appx_diagonalmua}
    \frac{\partial H_{j}}{\partial \mu_{{\rm a},j}} = & \frac{1}{A_j} \sum_{e} w_{e} \left( L_{e,i} \left(\exp\left[- \mu_{{\rm a},j} S_{{e},j}\right] - 1\right) \right. \\ & \left. +  S_{{e},j} \exp \left[-\mu_{{\rm a},i} S_{{e},j}\right]\right). 
  \end{split}
  \end{equation}
  
  For construction of the Jacobian for the scattering, perturbation
  Monte Carlo is utilized.  Then, the derivative with respect to the
  scattering coefficient can be computed using
  Eqs. \refeqn{eqn:appx_wmus2} and \refeqn{eqn:appx_abspert}.  An
  analytical expression for $\frac{\partial H_j}{\partial \mu_{{\rm
        s},i}}$ is obtained by writing the difference quotient and
  taking the limit ${\tilde{\mu}_{{\rm s},i} - \mu_{{\rm s},i}
    \rightarrow 0 }$
  \begin{equation}
  \begin{split}
  \label{eq:appx_limit_scattering_perturbation}
  \frac{\partial H_j}{\partial {\mu_{{\rm s},i}}} = &  
  \lim_{\tilde{\mu}_{{\rm s},i} - \mu_{{\rm s},i} \rightarrow 0 } 
  \frac{\tilde{H}_ j - H_j}{\tilde{\mu}_{{\rm s},i} - \mu_{{\rm s},i}} \\ = &
  \lim_{\Delta \mu_{{\rm s},i} \rightarrow 0 } 
  \frac{\Delta H_ j }{\Delta \mu_{{\rm s},i} } \\
  = &  \frac{1}{A_j}\lim_{\Delta \mu_{\rm s,i} \rightarrow 0 } \frac{1}{\Delta \mu_{{\rm s},i}}
  \left( \sum_{{e}} w_{e} \left(\frac{\mu_{{\rm s},i} + \Delta \mu_{{\rm s},i} }{\mu_{{\rm s},i}} \right )^{k_{{e},i}} \right. \\ & \cdot \exp\left[{-\Delta \mu_{{\rm s},i} L_{{e},i}}\right] \left(1-\exp\left[-\mu_{{\rm a},j} S_{e,j}\right]\right) \\ & \left. - \sum_{{e}} w_{e} \left(1 - \exp\left[{-\mu_{{\rm a},j} S_{{e},j}}\right]\right) \right)
  \end{split}
  \end{equation}
  The limit can be evaluated using the first order expansions for two the
  expressions in the numerator with $\Delta \mu_{{\rm s},i}$ to obtain
  \begin{equation}
  \begin{split}
    \label{eqn:_appx_musmua}
    \frac{\partial H_j}{\partial {\mu_{{\rm s},i}}}  = & \frac{1}{A_j} \sum_{e} w_{e} \left(\frac{k_{{e},i}}{\mu_{{\rm s},i}} - L_{{e},i} \right) \\ & \cdot \left(1 -\exp\left[-\mu_{{\rm a},j} S_{{e},j}\right]\right).
  \end{split}
  \end{equation}
  Computation of the Jacobian was validated with simulations described in Appendix \ref{sec:appx_validation_jacobian}.
  
  %================================================================================
  
  \section{Validation of the Jacobian computation}
  \label{sec:appx_validation_jacobian}
 
  Computation of the Jacobian was validated by evaluating derivatives of
  absorption and scattering, Eqs. \refeqn{eqn:nondiagonalmua},
  \refeqn{eqn:diagonalmua} and \refeqn{eqn:musmua}, and comparing them
  to a least-squares estimation of the derivatives from a finite
  difference approximation utilizing conventional Monte Carlo
  simulations.  A simple rectangular test geometry of size $3 \, {\rm
    mm} \times 3 \, {\rm mm}$ that was discretized into $81$ pixels was
  used.  The absorption and scattering coefficients of each pixel were
  selected at random in the intervals $\muaa \in [0 \, , \, 0.05] \,
  {\rm mm}^{-1}$ and $\muas \in [0 \, , \, 3] \, {\rm mm}^{-1}$.
  Scattering anisotropy was selected as a constant value for the whole
  simulation domain randomly in the interval of $g \in [-0.8 \, , \,
    0.8]$.
  
  The derivatives using the least squares approach \revold{that} were
  computed as follows. First two random pixels $i$ and $j$ were
  selected.  Symbol $i$ stands for the pixel in which the absorbed
  energy is evaluated and $j$ for the pixel where the optical
  coefficient is varied.  Then, to evaluate e.g. the derivative for
  absorption $J_{i,j} = \frac{\partial E_i}{\partial \mu_{{\rm a},j}}$,
  we introduced a small random variation $\delta \muaa$ to $\mu_{{\rm
      a},j}$ into the absorption coefficient of the pixel $j$ several
  times and computed the energy absorbed in the pixel $i$ using Monte
  Carlo.  Since $\delta \muaa$ is small, a least-squares line can be
  fitted to these values to estimate the derivative from the
  slope. Derivatives with respect to the scattering coefficient were
  evaluated in the same way.
  
\begin{figure}[!tp]
    \begin{center}
    \includegraphics[width=8cm]{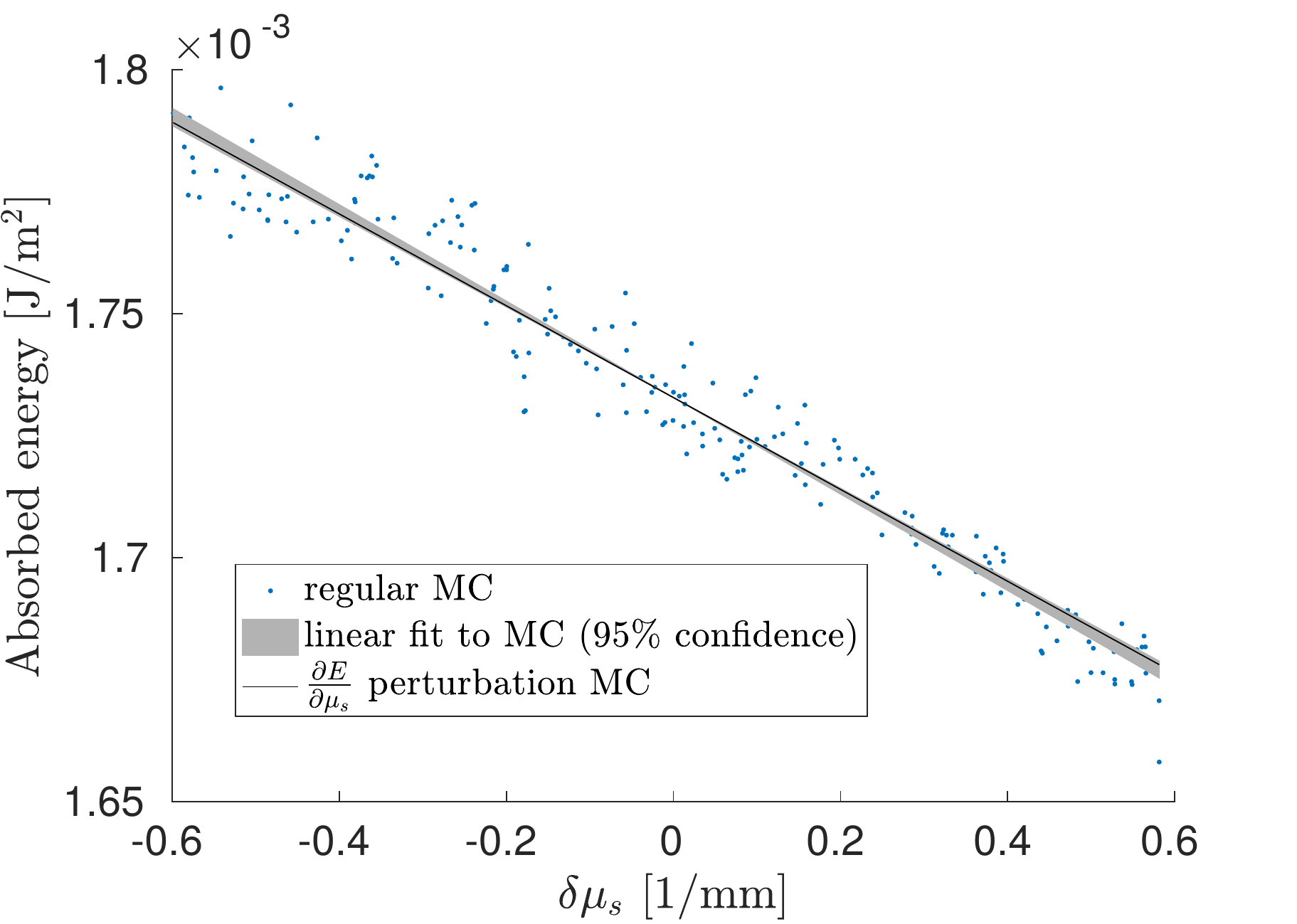}
    \end{center}
    \caption{ 
    Energy absorbed to one pixel of a rectangular simulation domain of size $3 \, {\rm mm} \times 3 \, {\rm mm}$ at a distance of $0.67 \, {\rm mm}$ from another pixel where scattering was varied, a linear fit for determining the slope
      $\frac{\partial E}{ \partial \muas}$ (grey region), and a line with a slope that is estimated at $\delta \muas = 0 $ using
      perturbation Monte Carlo (black line).} 
      \label{fig:validation}
  \end{figure}

  All the derivatives produced in this way agreed to those computed
  using Eqs. \refeqn{eqn:nondiagonalmua}, \refeqn{eqn:diagonalmua} and
  \refeqn{eqn:musmua}.  The photon count and the magnitude of the varied
  parameters were improved (i.e.  increasing photon count and decreasing
  the magnitude) until the results agreed within $95 \, \%$ confidence.
  However, the numerical noise in the data for the least-squares
  estimate sometimes became too high to obtain good enough relative
  accuracy for comparison.  Generally, this tends to happen if the
  distance between the pixels $i$ and $j$ is large.  An example of
  evaluating scattering derivatives using two pixels located at a
  distance of $0.67 \, {\rm mm}$ from each other is shown in
  Fig. \ref{fig:validation}.  In general, the simulations showed a good
  agreement between the derivatives evaluated using the two approaches.
  
\end{appendices}

%==================================================================================

\section*{Acknowledgment}

The authors would like to thank Professor Simon Arridge for valuable discussions.
This work was funded by the Academy of Finland (projects 314411 and 312342 Finnish Centre of Excellence in Inverse Modelling and Imaging) and 
Jane and Aatos Erkko foundation.


\begin{thebibliography}{10}
  \providecommand{\url}[1]{#1}
  \csname url@samestyle\endcsname
  \providecommand{\newblock}{\relax}
  \providecommand{\bibinfo}[2]{#2}
  \providecommand{\BIBentrySTDinterwordspacing}{\spaceskip=0pt\relax}
  \providecommand{\BIBentryALTinterwordstretchfactor}{4}
  \providecommand{\BIBentryALTinterwordspacing}{\spaceskip=\fontdimen2\font plus
  \BIBentryALTinterwordstretchfactor\fontdimen3\font minus
    \fontdimen4\font\relax}
  \providecommand{\BIBforeignlanguage}[2]{{%
  \expandafter\ifx\csname l@#1\endcsname\relax
  \typeout{** WARNING: IEEEtran.bst: No hyphenation pattern has been}%
  \typeout{** loaded for the language `#1'. Using the pattern for}%
  \typeout{** the default language instead.}%
  \else
  \language=\csname l@#1\endcsname
  \fi
  #2}}
  \providecommand{\BIBdecl}{\relax}
  \BIBdecl
  
  \bibitem{xu2006}
  M.~Xu and L.~V. Wang, ``Photoacoustic imaging in biomedicine,'' \emph{Rev. Sci.
    Instrum.}, vol.~77, p. 041101, 2006.
  
  \bibitem{li2009}
  C.~Li and L.~V. Wang, ``Photoacoustic tomography and sensing in biomedicine,''
    \emph{Phys. Med. Biol.}, vol.~54, pp. R59--R97, 2009.
  
  \bibitem{wang2009}
  L.~V. Wang, Ed., \emph{Photoacoustic Imaging and Spectroscopy}.\hskip 1em plus
    0.5em minus 0.4em\relax CRC Press, 2009.
  
  \bibitem{beard2011}
  P.~Beard, ``Biomedical photoacoustic imaging,'' \emph{Interface Focus}, vol.~1,
    pp. 602--631, 2011.
  
  \bibitem{xia2014}
  J.~Xia and L.~V. Wang, ``Small-animal whole-body photoacoustic tomography: a
    review,'' \emph{Phys. Med. Biol.}, vol.~61, no.~5, pp. 1380--1389, 2014.
  
  \bibitem{wang2016}
  L.~V. Wang and J.~Yao, ``A practical guide to photoacoustic tomography in the
    life sciences,'' \emph{Nat. Methods}, vol.~13, no.~8, pp. 627--638, 2016.
  
  \bibitem{weber2016}
  J.~Weber, P.~C. Beard, and S.~Bohndiek, ``Contrast agents for molecular
    photoacoustic imaging,'' \emph{Nat. Methods}, vol.~13, no.~8, pp. 639--650,
    2016.
  
  \bibitem{brunker2017}
  J.~Brunker, J.~Yao, J.~Laufer, and S.~E. Bohndiek, ``Photoacoustic imaging
    using genetically encoded reporters: a review,'' \emph{J. Biomed. Opt.},
    vol.~22, no.~7, p. 070901, 2017.
  
  \bibitem{cox2012a}
  B.~Cox, J.~G. Laufer, S.~R. Arridge, and P.~C. Beard, ``Quantitative
    spectroscopic photoacoustic imaging{:} a review,'' \emph{J. Biomed. Opt.},
    vol.~17, no.~6, p. 061202, 2012.
  
  \bibitem{ishimaru78a}
  A.~Ishimaru, \emph{Wave Propagation and Scattering in Random Media}.\hskip 1em
    plus 0.5em minus 0.4em\relax New York: Academic Press, 1978, vol.~1.
  
  \bibitem{arridge99}
  S.~R. Arridge, ``Optical tomography in medical imaging,'' \emph{Inv. Probl.},
    vol.~15, pp. R41--R93, 1999.
  
  \bibitem{prahl89}
  S.~A. Prahl, M.~Keijzer, S.~L. Jacques, and A.~J. Welch, ``A {M}onte {C}arlo
    model of light propagation in tissue,'' in \emph{Proc. SPIE, 
    Dosimetry of Laser Radiation in Medicine and Biology}, G.~M\"uller and
    D.~Sliney, Eds., vol. IS 5, 1989, pp. 102--111.
  
  \bibitem{wang1995}
  L.~Wang, S.~Jacques, and L.~Zheng, ``{MCML} {--} {M}onte {C}arlo modeling of
    photon transport in multi-layered tissues,'' \emph{Comput. Methods Programs
    Biomed.}, vol.~47, pp. 131--146, 1995.
  
  \bibitem{sassaroli2012}
  A.~Sassaroli and F.~Martelli, ``Equivalence of four {M}onte {C}arlo methods for
    photon migration in turbid media,'' \emph{J. Opt. Soc. Am. A}, vol.~29, no.~10,
    pp. 2110--2117, 2012.
  
  \bibitem{zhu2013}
  C.~Zhu and Q.~Liu, ``Review of {M}onte {C}arlo modeling of light transport in
    tissues,'' \emph{J. Biomed. Opt.}, vol.~18, no.~5, p. 050902, 2013.
  
  \bibitem{hayakawa2014}
  C.~Hayakawa, J.~Spanier, and V.~Venugopalan, ``Comparative analysis of discrete
    and continuous absorption weighting estimators used in {M}onte {C}arlo
    simulations of radiative transport in turbid media,'' \emph{J. Opt. Soc. Am. A},
    vol.~31, no.~2, pp. 301--311, 2014.
  
  \bibitem{fang2009}
  Q.~Fang and D.~Boas, ``{M}onte {C}arlo simulation of photon migration in {3D}
    turbid media accelerated by graphics processing units,'' \emph{Opt. Express},
    vol.~17, no.~22, pp. 20\,178--20\,190, 2009.
  
  \bibitem{yang2013}
  O.~Yang and B.~Choi, ``Accelerated rescaling of single {M}onte {C}arlo
    simulation runs with the {G}raphics {P}rocessing {U}nit {(GPU)},''
    \emph{Biomed. Opt. Express}, vol.~4, no.~11, pp. 2667--2672, 2013.
  
  \bibitem{cassidy2018}
  J.~Cassidy, A.~Nouri, V.~Betz, and L.~Lilge, ``High-performance, robustly
    verified {Monte Carlo} simulation with {FullMonte},'' \emph{J.
    Biomed. Opt.}, vol.~23, no.~8, p. 085001, 2018.
  
  \bibitem{liu2015}
  Y.~Liu, S.~Jacques, M.~Azimipour, J.~Rogers, R.~Pashaie, and K.~Eliceiri,
    ``{OptogenSIM}: a {3D} {M}onte {C}arlo simulation platform for light delivery
    design in optogenetics,'' \emph{Biomed. Opt. Express}, vol.~6, no.~12, pp.
    4859--4870, 2015.
  
  \bibitem{Leino19}
  A.~A. Leino, A.~Pulkkinen, and T.~Tarvainen, ``Valo{MC}: a {M}onte {C}arlo
    software and {MATLAB} toolbox for simulating light transport in biological
    tissue,'' \emph{OSA Continuum}, vol.~2, no.~3, pp. 957--972, Mar 2019.
  
  \bibitem{tarvainen2012}
  T.~Tarvainen, B.~T. Cox, J.~P. Kaipio, and S.~R. Arridge, ``Reconstructing
    absorption and scattering distributions in quantitative photoacoustic
    tomography,'' \emph{Inv. Probl.}, vol.~28, p. 084009, 2012.
  
  \bibitem{saratoon2013}
  T.~Saratoon, T.~Tarvainen, B.~T. Cox, and S.~R. Arridge, ``A gradient-based
    method for quantitative photoacoustic tomography using the radiative transfer
    equation,'' \emph{Inv. Probl.}, vol.~29, p. 075006, 2013.
  
  \bibitem{mamonov2014}
  A.~V. Mamonov and K.~Ren, ``Quantitative photoacoustic imaging in radiative
    transport regime,'' \emph{Comm. Math. Sci.}, vol.~12, pp. 201--234, 2014.
  
  \bibitem{haltmeier2015}
  M.~Haltmeier, L.~Neumann, and S.~Rabanser, ``Single-stage reconstruction
    algorithm for quantitative photoacoustic tomography,'' \emph{Inv. Probl.},
    vol.~31, p. 065005, 2015.
  
  \bibitem{buchmann2017}
  J.~Buchmann, B.~A. Kaplan, S.~Prohaska, and J.~Laufer, ``Experimental
    validation of a {M}onte-{C}arlo-based inversion scheme for {3D} quantitative
    photoacoustic tomography,'' in \emph{Photons Plus Ultrasound: Imaging and
    Sensing 2017, Proc of SPIE}, A.~Oraevsky and L.~Wang, Eds., vol. 10064, 2017,
    p. 1006416.
  
  \bibitem{kaplan2017}
  B.~A. Kaplan, J.~Buchmann, S.~Prohaska, and J.~Laufer, ``Monte-{C}arlo-based
    inversion scheme for {3D} quantitative photoacoustic tomography,'' in
    \emph{Photons Plus Ultrasound: Imaging and Sensing 2017, Proc of SPIE},
    A.~Oraevsky and L.~Wang, Eds., vol. 10064, 2017, p. 100645J.
  
  \bibitem{hochuli2016}
  R.~Hochuli, S.~Powell, S.~Arridge, and B.~Cox, ``Quantitative photoacoustic
    tomography using forward and adjoint {M}onte {C}arlo models of radiance,''
    \emph{J Biomed Opt}, vol.~21, no.~12, p. 126004, 2016.
  
  \bibitem{Buchmann2019}
  J.~Buchmann, B.~A. Kaplan, S.~Powell, S.~Prohaska, and J.~Laufer,
    ``Three-dimensional quantitative photoacoustic tomography using an adjoint
    radiance {M}onte {C}arlo model and gradient descent,'' \emph{J. Biomed. Opt.},
    vol.~24, no.~6, p. 066001, 2019.
    
      \bibitem{buchmann2020} 
     {J.~Buchmann, B.~Kaplan, S.~Powell, S.~Prohaska, and J.~Laufer, ``Quantitative {PA} tomography of high resolution {3-D} images: experimental validation in a tissue phantom,'' \emph{Photoacoustics}, vol.~17, p. 100157, 2020.}
  
  \bibitem{Ivan1991}
  I.~Lux and L.~Koblinger, \emph{Monte Carlo Particle Transport Methods: Neutron
    and Photon Calculations}.\hskip 1em plus 0.5em minus 0.4em\relax CRC Press,
    1991.
  
 \bibitem{kienle1996}
  {A.~Kienle and M.~S.~Patterson, ``Determination of the optical properties of turbid media
from a single {M}onte {C}arlo simulation,'' \emph{Phys.
    Med. Biol.}, vol.~41, pp. 2221-2227, 1996.}
    
   \bibitem{pifferi1998}
  {A.~Pifferi, P.~Taroni, G.~Valentini, and S.~Andersson-Engels, ``Real-time method for fitting time-resolved reflectance and transmittance measurements with a {M}onte {C}arlo model,'' \emph{Appl.
    Opt.}, vol.~37, no.~13, pp. 2774-2780, 1998.}
    
  \bibitem{Hayakawa2001}
  C.~K. Hayakawa, J.~Spanier, F.~Bevilacqua, A.~K. Dunn, J.~S. You, B.~J.
    Tromberg et al., ``Perturbation {M}onte {C}arlo methods to solve
    inverse photon migration problems in heterogeneous tissues,'' \emph{Opt.
    Lett.}, vol.~26, no.~17, pp. 1335--1337, Sep 2001.
  
  \bibitem{kumar2004}
  Y.~Kumar and R.~Vasu, ``Reconstruction of optical properties of low-scattering
    tissue using derivative estimated through perturbation {M}onte-{C}arlo
    method,'' \emph{J. Biomed. Opt.}, vol.~9, no.~5, pp. 1002--1012, 2004.

\bibitem{alerstam2008}
  {E.~Alerstam, S.~Andersson-Engels, and T.~Svensson ``White {M}onte {C}arlo for time-resolved photon migration,'' \emph{J. Biomed.
    Opt.}, vol.~13, no.~4, p. 041304, 2008.}  
  
  \bibitem{chen2009}
  J.~Chen and X.~Intes, ``Time-gated perturbation {M}onte {C}arlo for whole body
    functional imaging in small animals,'' \emph{Opt. Express.}, vol.~17, no.~22,
    pp. 19\,566--19\,579, 2009.
  
  \bibitem{sassaroli2011}
  A.~Sassaroli, ``Fast perturbation {M}onte {C}arlo method for photon migration
    in heterogeneous turbid media,'' \emph{Opt. Lett.}, vol.~36, no.~11, pp.
    2095--2097, 2011.
  
  \bibitem{zhang2012}
  X.~Zhang, ``Construction of the {J}acobian matrix for fluorescence diffuse
    optical tomography using a perturbation {M}onte {C}arlo method,'' in
    \emph{Multimodal Biomedical Imaging VII, Proc. of SPIE}, vol. 8216, 2012, p.
    82160O.
  
  \bibitem{yamamoto2016}
  T.~Yamamoto and H.~Sakamoto, ``Frequency domain optical tomography using a
    {M}onte {C}arlo perturbation method,'' \emph{Opt. Comm.}, vol. 364, pp.
    165--176, 2016.
  
  \bibitem{yao2018}
  R.~Yao, X.~Intes, and Q.~Fang, ``Direct approach to compute {J}acobians for
    diffuse optical tomography using perturbation {M}onte {C}arlo-based photon
    "replay",'' \emph{Biomed. Opt. Express}, vol.~9, no.~10, pp. 4588--4603, 2018.
  
  \bibitem{henyey41}
  L.~G. Henyey and J.~L. Greenstein, ``Diffuse radiation in the galaxy,''
    \emph{AstroPhys. J.}, vol.~93, pp. 70--83, 1941.
  
  \bibitem{cox2005}
  B.~T. Cox and P.~C. Beard, ``Fast calculation of pulsed photoacoustic fields in
    fluids using {\emph{k}-}space methods,'' \emph{J. Acoust. Soc. Am.}, vol. 117,
    no.~6, pp. 3616--3627, 2005.
  
  \bibitem{treeby2010a}
  B.~E. Treeby and B.~T. Cox, ``{k}{-}{W}ave{:} {M}{A}{T}{L}{A}{B} toolbox for
    the simulation and reconstruction of photoacoustic wave fields,'' \emph{J.
    Biomed, Opt.}, vol.~15, no.~2, p. 021314, 2010.
  
  \bibitem{kaipio05}
  J.~Kaipio and E.~Somersalo, \emph{Statistical and Computational Inverse
    Problems}.\hskip 1em plus 0.5em minus 0.4em\relax New York: Springer, 2005.
  
  \bibitem{tarvainen2013}
  T.~Tarvainen, A.~Pulkkinen, B.~T. Cox, J.~P. Kaipio, and S.~R. Arridge,
    ``Bayesian image reconstruction in quantitative photoacoustic tomography,''
    \emph{IEEE Trans. Med. Imag.}, vol.~32, no.~12, pp. 2287--2298, 2013.
  
  \bibitem{rasmussen2006}
  C.~E. Rasmussen and C.~K.~I. Williams, \emph{Gaussian Processes for Machine
    Learning}.\hskip 1em plus 0.5em minus 0.4em\relax MIT Press, 2006.
  
  \bibitem{pulkkinen2014}
  A.~Pulkkinen, B.~T. Cox, S.~R. Arridge, J.~P. Kaipio, and T.~Tarvainen, ``A
    {B}ayesian approach to spectral quantitative photoacoustic tomography,''
    \emph{Inv. Probl.}, vol.~30, p. 065012, 2014.
  
  \bibitem{Nocedal06}
  J.~Nocedal and S.~Wright, \emph{Numerical Optimization}.\hskip 1em plus 0.5em
    minus 0.4em\relax Springer, 2006.
  
  \bibitem{jacques2013}
  S.~Jacques, ``Optical properties of biological tissues: a review,'' \emph{Phys.
    Med. Biol.}, vol.~58, no.~11, pp. R37--R61, 2013.
  
  \bibitem{friebel2006}
  M.~Friebel, A.~Roggan, G.~M{\"u}ller, and M.~Meinke, ``Determination of optical
    properties of human blood in the spectral range 250 to 1100 nm using {M}onte
    {C}arlo simulations with hematocrit-dependent effective scattering phase
    functions,'' \emph{J Biomed. Opt.}, vol.~11, no.~3, p. 034021, 2006.
  
  \bibitem{schoberl1997netgen}
  J.~Sch{\"o}berl, ``{NETGEN} {A}n advancing front {2D/3D}-mesh generator based
    on abstract rules,'' \emph{Comput. Vis. Sci.}, vol.~1, no.~1, pp. 41--52, 1997.
  
  \bibitem{zhang2008a}
  E.~Zhang, J.~Laufer, and P.~Beard, ``Backward-mode multiwavelength
    photoacoustic scanner using a planar {F}abry-{P}erot polymer film ultrasound
    sensor for high-resolution three-dimensional imaging of biological tissues,''
    \emph{Appl. Opt.}, vol.~47, no.~4, pp. 561--577, 2008.

\bibitem{buchmann2016}
{J.~Buchmann, E.~Zhang, C.~Scharfenorth, B.~Spannekrebs, C.~Villringer, and J.~Laufer, "Evaluation of Fabry-Perot
  polymer film sensors made using hard dielectric mirror deposition", in
  \emph{Photons Plus Ultrasound: Imaging and
    Sensing 2017, Proc of SPIE}, A.~Oraevsky and L.~Wang, Eds., vol. 9708, 2016, p. 970856.}
  
  \bibitem{ellwood2017}
  R.~Ellwood, O.~Ogunlade, E.~Zhang, P.~Beard, and B.~Cox, ``Photoacoustic
    tomography using orthogonal {F}abry-{P}{\'e}rot sensors,'' \emph{J. Biomed.
    Opt.}, vol.~22, no.~4, p. 041009, 2017.
  
  
  
  
  \end{thebibliography}
\end{document}

\title{Perturbation Monte Carlo Method for \\ Quantitative Photoacoustic Tomography -- \\ Supplementary Material}

\author[a]{Aleksi Leino}
\author[a]{Tuomas Lunttila}
\author[a]{Meghdoot Mozumder}
\author[a]{Aki Pulkkinen}
\author[a,b]{Tanja Tarvainen}

\affil[a]{\small Department of Applied Physics, University of Eastern Finland, P.O. Box 1627, 70211 Kuopio, Finland}
\affil[b]{Department of Computer Science, University College London, Gower Street, WC1E 6BT, London, United Kingdom}

\maketitle

\section{Comparison with Adjoint Monte Carlo Simulations}

\begin{figure}[tp!]
  \begin{center}
  \includegraphics[width=13cm]{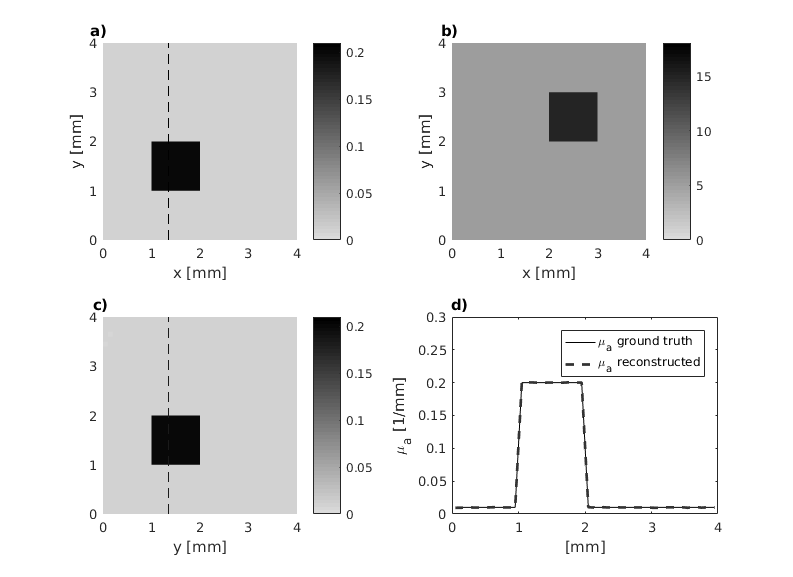}
  \end{center}
  \caption{a) The absorption coefficient to be reconstructed. b) The
    known scattering coefficient. c) The reconstructed absorption
    coefficient under the assumption that the scattering is known. d)
    Values of optical absorption along the lines shown in a) and c).}
    \label{fig:mua}
\end{figure}

\begin{figure}[tp!]
  \begin{center}
  \includegraphics[width=13cm]{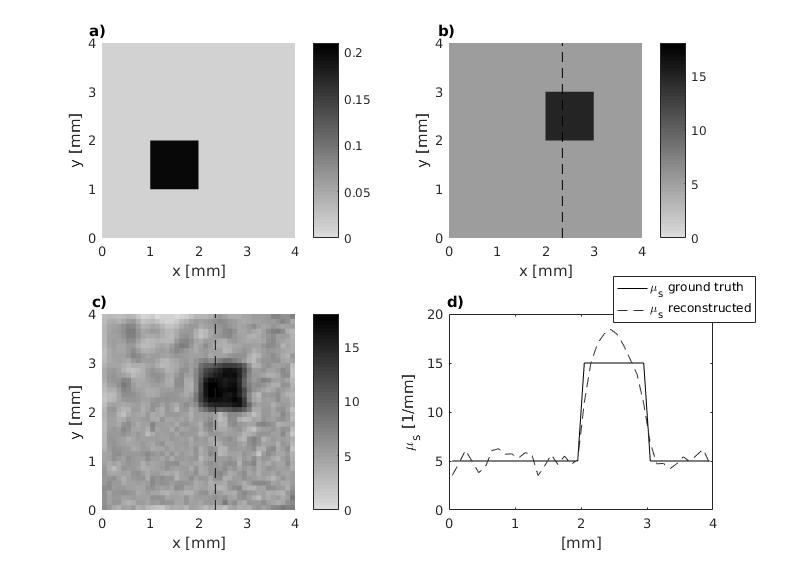}
  \end{center}
  \caption{a) The known absorption coefficient. b) The optical
    scattering to be reconstructed. c) The reconstructed scattering
    coefficient under the assumption that the absorption is known. d)
    Values of optical scattering along the lines shown in a) and c).}
    \label{fig:mus}
\end{figure}

To compare perturbation Monte Carlo (PMC) to the adjoint Monte Carlo method we
performed two simulations following
the simulation setup presented in Refs. \cite{hochuli2016,hochuli2016b}. 
In this setup, a $4 \: \mathrm{mm} \times 4 \: \mathrm{mm}$
domain with square inclusions embedded in a background with constant optical parameters was considered. 
The absorption and scattering coefficients of the background were $\muaa = 0.01 \, {\rm mm}^{-1}$ and $\muas = 5 \, {\rm mm}^{-1}$, and of the inclusions $\muaa = 0.2 \, {\rm mm}^{-1}$ and $\muas = 15 \, {\rm mm}^{-1}$.
The anisotropy parameter of the Henyey-Greenstein phase function was $g=0.9$. 
The simulation geometry and the parameters are illustrated in Figs. \ref{fig:mua} and \ref{fig:mus}.

For data simulation, the domain was discretized in a $101 \times 101$ pixel grid, where the pixels were split into  $20402$ triangular elements with $10404$ nodes.  
Photon fluence and absorbed optical energy density were simulated with the Monte Carlo method implemented with
ValoMC MATLAB toolbox \cite{Leino19}. 
The target was illuminated from the left and  top faces separately with a
collimated light source with the whole face acting
as a light source.
The number of photon packets used for each illumination was $10^8$.
Gaussian noise with standard deviation equal to  $0.1 \%$ of the maximum value of the
simulated data was added to the data.

In the solution of the inverse problem, the photon fluence and absorbed optical energy density were represented in a regular grid (two triangles form a rectangle) composed of  $3200$ elements and  $1681$ nodes. 
The absorption and scattering parameters were represented in a
$40 \times 40$ rectangular pixel grid. 
Reconstructions of optical absorption and scattering were performed separately assuming the other parameter to be known.
The measurement noise statistics was assumed to be known.
For the prior we used Ornstein-Uhlenbeck prior where the expected value of the prior was set to the background optical parameter values and the standard deviation was set such that the maximum
target values corresponded to two standard deviations
from the expected value.
The characteristic length scale was set as $0.5 \, {\rm mm}$.
The MAP estimates were computed using the Gauss-Newton method. As the initial guess for absorption and scattering in the iterations, the expected values of the prior were used. 
During the iterations, the forward solutions and the Jacobians were computed using Monte Carlo and PMC using $10^8$ photon packets for each illumination.

Fig. \ref{fig:mua} shows the results of estimating the optical
absorption when the scattering is known.
Convergence was achieved in 4 Gauss-Newton iterations.  
The relative error of the estimated absorption was $0.7\%$, in comparison to $0.2 \%$ reported in
Ref. \cite{hochuli2016}. 
Fig. \ref{fig:mus} shows the results of estimating the optical
scattering when the absorption is known.
Convergence was achieved in 6 Gauss-Newton iterations.  
The relative error of the estimated scattering was $22\%$. 
Authors in Ref. \cite{hochuli2016} did not
report relative errors for the estimated scattering. 
Visual comparison to their results in estimation of scattering favors results
obtained using the presented approach. 
However, the authors of Ref. \cite{hochuli2016} did not use any regularization in their approach. 

Perturbation Monte Carlo performs well in terms of both visual quality of the
reconstructions and absolute values of relative errors.  
One of the appealing features of the PMC in comparison to the adjoint approach is that
there is no need to store the angular solution. 
This eases the memory consumption and avoids introducing a new parameter to be discretized. 
With our current hardware, model and implementation, most memory is
consumed in storing the Jacobians.  
In the current PMC implementation, each thread reserves an array for $L_{e,i}$ (total length propagated
in region $i$) and $k_{e,i}$ (total number of scattering events in
region $i$). 
There is no need to store the photon paths. 
The memory consumption could become non-trivial on a highly parallelized platform
using a three-dimensional model.